\documentclass[aps,prl,twocolumn,nofootinbib,nobibnotes]{revtex4-1}
\usepackage[utf8]{inputenc}
\usepackage{amssymb}
\usepackage{amsmath}
\usepackage{amsfonts}
\usepackage{graphicx}
\usepackage{color}
\usepackage{xspace}
\usepackage{comment}
\usepackage[hypertexnames=false]{hyperref}
\usepackage[normalem]{ulem}

\usepackage[section]{placeins}
\usepackage{afterpage}

\usepackage{float}
\usepackage{slashed}
\usepackage{appendix}

\usepackage{cancel}

\usepackage{multirow,rotating}
\usepackage[dvipsnames]{xcolor}
\usepackage{orcidlink}   

\ifdefined\pdfminorversion\pdfminorversion=7\fi
\begin{document}
	
      \title{Vertexing displaced diphoton decays with recoil photons at Belle II}
      
	   \author{Zeren Simon Wang\,\orcidlink{0000-0002-1483-6314}}
	   \email{wzs@hfut.edu.cn}
      \affiliation{School of Physics, Hefei University of Technology, Hefei 230601, People’s Republic of China}

      \author{Yu Zhang\,\orcidlink{0000-0001-9415-8252}}
      \email{dayu@hfut.edu.cn}
      \affiliation{School of Physics, Hefei University of Technology, Hefei 230601, People’s Republic of China}

\begin{abstract}
We propose a recoil-assisted strategy for vertexing displaced diphoton decays in $e^+e^-\to\gamma X$, $X\to\gamma\gamma$, using only one converted daughter photon. The initial state and recoil-photon momentum define the LLP flight line, whose closest approach to the converted-photon trajectory locates the decay vertex. This removes the need for a second conversion, making the conversion-related efficiency scale linearly rather than quadratically with the photon-conversion probability. For photophilic axionlike particles at Belle II, the existing $408$ fb$^{-1}$ data set could probe previously unconstrained parameter space near $m_a \simeq 100$ MeV and $g_{a\gamma\gamma} \simeq 2 \times 10^{-4}$ GeV$^{-1}$, improving leading bounds by almost an order of magnitude and testing the mass region below the $\pi^0$ mass, which the existing Belle II three-photon searches cannot access. With $50$ ab$^{-1}$, the reach extends to $m_a \simeq 310$ MeV and $g_{a\gamma\gamma}\simeq2\times10^{-5}$ GeV$^{-1}$.
\end{abstract}
		

	\maketitle
    \noindent

\emph{\textbf{Introduction}---}
Searches for long-lived particles (LLPs) probe physics beyond the Standard Model (BSM) in which feeble interactions or suppressed decay modes allow particles to travel macroscopic distances~\cite{Lee:2018pag,Alimena:2019zri,Jeanty:2025wai}.
At colliders, LLPs can give rise to displaced vertices, disappearing tracks, displaced jets or leptons, and non-pointing photons, depending on their lifetime, charge, decay position, and final state.
Displaced diphoton decays, which arise in BSM scenarios including photophilic axionlike particles (ALPs), are particularly difficult to reconstruct because photons leave no charged tracks at the decay point.
Existing searches exploit calorimeter pointing and timing to recover directional or decay-position information.
ATLAS has searched for non-pointing and delayed photons using the longitudinal segmentation and timing of its electromagnetic calorimeter~\cite{ATLAS:2013etx,ATLAS:2014kbb}, while CMS has used precision calorimeter timing in delayed-photon searches~\cite{CMS:2019zxa}.

For displaced diphoton decays, reconstructing a three-dimensional displaced vertex (DV) generally requires an independent direction measurement for each photon or additional event-level constraints.
ATLAS recently used two photon trajectories inferred from calorimeter measurements to locate a common displaced diphoton vertex in the $(R,z)$ plane~\cite{ATLAS:2023meo}.
Photon conversion into an electron-positron pair inside the tracking system provides a full three-dimensional trajectory, and a recent phenomenological study proposed reconstructing displaced $a\to\gamma\gamma$ vertices at ATLAS from the intersection of two converted-photon trajectories~\cite{Alonso-Alvarez:2023wni}.
Requiring both photons to convert, however, makes the conversion-related efficiency quadratic in the per-photon conversion probability.
This strongly reduces the efficiency because that probability is only at the percent level in the relevant low-material detector volume.

\begin{figure}[t]
\centering
\includegraphics[width=0.99\linewidth]{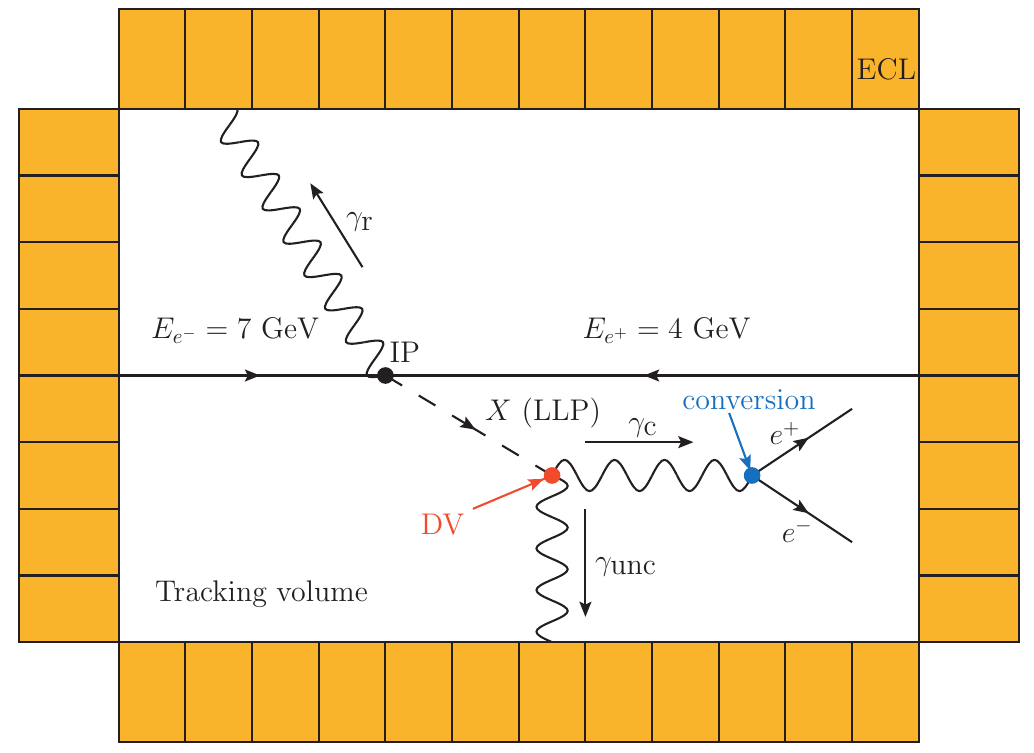}
\caption{Event topology and recoil-assisted search strategy at Belle~II. $\gamma_{\text{r}}$, $\gamma_{\text{unc}}$, and $\gamma_{\text{c}}$ denote the recoil, unconverted, and converted photons, respectively.}
\label{fig:event_topology}
\end{figure}

We propose a recoil-assisted strategy that uses electron-positron collision kinematics to replace one of the two daughter-photon directional measurements.
For the process
\begin{equation}
e^+e^-\to\gamma_r X,\qquad X\to\gamma\gamma,
\label{eqn:signal_topology}
\end{equation}
where $X$ is a neutral LLP, we determine the four-momentum $k_{\gamma_r}$ of the prompt recoil photon $\gamma_r$ from its measured energy and position.
The known initial-state four-momentum $P_{\rm ini}$ then fixes the LLP momentum, $p_X=P_{\rm ini}-k_{\gamma_r}$, and hence its flight line from the interaction point (IP).
If either daughter photon converts, the conversion point and the combined momentum of the charged tracks define its trajectory.
We estimate the DV from the closest approach of this trajectory to the LLP flight line, as illustrated in Fig.~\ref{fig:event_topology}.
The conversion-related efficiency therefore scales linearly with the photon-conversion probability.

This reconstruction relies on the known initial state at electron-positron colliders; in inclusive proton-proton collisions, the unknown incoming-parton momentum fractions generally prevent a recoil photon alone from determining the LLP four-momentum and three-dimensional flight line.
Related ingredients have previously been used separately: KLOE used recoil kinematics together with calorimeter timing to reconstruct $K_L\to3\pi^0$ decay vertices~\cite{KLOE:2005lau}, while CMS used conversion tracks to measure the transverse impact parameter of non-pointing photons~\cite{CMS:2012idp}.
To the best of our knowledge, this work proposes the first recoil-assisted single-conversion strategy for three-dimensional vertexing of displaced diphoton decays at electron-positron colliders.

We use a photophilic ALP as a benchmark realization of $X$.
ALPs are generic pseudo-Nambu--Goldstone bosons and can naturally be light and weakly coupled~\cite{Bauer:2017ris,Bauer:2018uxu}.
A KSVZ-like model with a colorless, electrically charged PQ-chiral fermion gives a simple photophilic realization: the fermion generates an electromagnetic anomaly but no QCD anomaly~\cite{Dias:2014osa}.
The same dimension-five operator then governs both radiative ALP production and its diphoton decay.
For sub-GeV masses, the cubic mass dependence of the decay width, combined with the large boost at electron-positron colliders, naturally gives detector-scale decay lengths.
At Belle~II, prompt ALP decays lead to a three-photon final state, whereas sufficiently long-lived ALPs escape the detector and appear as missing energy.
Earlier projections identified an intermediate-lifetime gap between these two regimes and noted that a dedicated displaced-photon search could extend the sensitivity there~\cite{Dolan:2017osp}.
Recoil-assisted vertexing directly targets this gap.

Belle~II is therefore a natural benchmark: its electromagnetic calorimetry, charged-particle tracking, and vertex reconstruction are well suited to this search.
Our Monte Carlo (MC) study uses a parameterized detector response, includes the leading irreducible multiphoton QED backgrounds, and evaluates the reach of the existing $408~\text{fb}^{-1}$ and projected $50~\text{ab}^{-1}$ data sets.
We find that the existing $408~\text{fb}^{-1}$ data set could exclude ALP-photon couplings down to $g_{a\gamma\gamma} \simeq 2\times 10^{-4}~\text{GeV}^{-1}$ near $m_a \simeq 100$~MeV at $95\%$ confidence level (CL).
This would improve the leading existing constraints by almost one order of magnitude in the most sensitive region and probe large lifetimes and masses below the $\pi^0$ threshold which the recent Belle~II prompt three-photon search~\cite{Belle-II:2026znr} cannot access.
With $50~\text{ab}^{-1}$, the reach extends to couplings smaller by roughly another order of magnitude in the most favorable mass range and to masses up to $m_a \simeq 310$~MeV.

More broadly, the method applies to other neutral LLPs produced with a photon and decaying into two photons, including massive spin-2 states~\cite{dEnterria:2023npy} and \textit{CP}-even scalars~\cite{Perazzi:2000id,Pustyntsev:2024ygw}.
The strategy can also be adapted to other electron-positron colliders, including BESIII~\cite{BESIII:2009fln} and proposed future facilities such as the Super Tau-Charm Facility (STCF)~\cite{Achasov:2023gey,Ai:2025xop} and FCC-ee~\cite{FCC:2018evy}.

\vspace{0.2cm}
\emph{\textbf{Photophilic ALPs at Belle~II}---}
We describe the photophilic ALP interaction with the Lagrangian
\begin{eqnarray}
    \mathcal{L}_a= - \frac{g_{a\gamma\gamma}}{4} \, a F_{\mu\nu}\widetilde{F}^{\mu\nu},
\end{eqnarray}
where $a$ denotes the ALP field, $F_{\mu\nu}$ is the electromagnetic field-strength tensor, $\widetilde{F}^{\mu\nu}\equiv\epsilon^{\mu\nu\rho\sigma}F_{\rho\sigma}/2$ is its dual, and $g_{a\gamma\gamma}$ is a coupling of mass dimension $[-1]$.

For unpolarized beams and negligible electron mass, the tree-level cross section for radiative ALP production is~\cite{Bauer:2018uxu}
\begin{equation}
\sigma(e^+e^-\to a\gamma)=\frac{\alpha(s)\,g_{a\gamma\gamma}^2}{24}
\left(1-\frac{m_a^2}{s}\right)^3,
\label{eqn:signal_XS}
\end{equation}
where $\sqrt{s}$ is the center-of-mass (COM) energy and $\alpha(s)$ is the electromagnetic fine-structure constant at this scale.
We assume $m_a<\sqrt{s}$.
The corresponding diphoton decay width is
\begin{equation}
\Gamma(a\to\gamma\gamma)=\frac{g_{a\gamma\gamma}^2m_a^3}{64\pi}.
\label{eqn:decaywidth}
\end{equation}
At leading order in the minimal photophilic scenario, this is the only relevant ALP decay mode, so $\text{BR}(a\to\gamma\gamma) \simeq 1$.
The same coupling controls both the production rate and the decay length.

SuperKEKB~\cite{Ohnishi:2013fma,Akai:2018mbz}, where Belle~II operates, collides a 7-GeV electron beam with a 4-GeV positron beam at $\sqrt{s}=10.58$~GeV near the $\Upsilon(4S)$ resonance.
For $m_a\ll\sqrt{s}$, we boost the two-body kinematics to the laboratory frame and average over the COM-frame production angle to obtain the mean decay length
\begin{equation}
    \left\langle\lambda_a^{\text{lab}}\right\rangle \simeq 218~\text{mm}     \left(\frac{10^{-4}~\text{GeV}^{-1}}{g_{a\gamma\gamma}}\right)^2
    \left(\frac{100~\text{MeV}}{m_a}\right)^4.
    \label{eqn:decayLength_lab_estimate}
\end{equation}
Thus, the region around $(m_a,g_{a\gamma\gamma})\sim(100~\text{MeV},10^{-4}~\text{GeV}^{-1})$ gives detector-scale decay lengths suited to this search at Belle~II.
The Supplemental Material gives the production-angle distribution, two-body COM kinematics, boost transformation, and general expression for the angle-averaged decay length.

\vspace{0.2cm}
\emph{\textbf{Recoil-assisted single-conversion search}---}
We simulate $e^+e^-\to a\gamma$, $a\to\gamma\gamma$ with a parameterized detector response guided by published Belle~II studies~\cite{Belle-II:2018jsg}.
We generate hard events with \texttt{MadGraph5\_aMC\_v3.6.6}~\cite{Alwall:2014hca,Frederix:2018nkq}.
We include leading-logarithmic QED initial-state radiation (ISR) through electron and positron structure functions and an exclusive lepton-QED shower in \texttt{Pythia~8.3}~\cite{Sjostrand:2014zea,Bierlich:2022pfr}.

The target topology contains an unconverted prompt recoil photon, exactly one converted ALP daughter, and an unconverted daughter reaching the electromagnetic calorimeter (ECL).
We calculate photon-conversion and survival probabilities from an effective material-budget model of the beam pipe, two pixel-detector (PXD) layers, and four silicon-vertex-detector (SVD) layers~\cite{Belle-II:2010dht}.
We use the Run~1 vertex-detector (VXD) configuration for $408$~fb$^{-1}$ and the full VXD for $50$~ab$^{-1}$.
These configurations differ only in the modeled PXD~L2 coverage, which is partial in Run~1 and complete in the full VXD~\cite{Belle-IIDEPFET:2024iqz}.

We smear the converted-photon momentum and conversion vertex directly.
For both signal and background, we take a constant conditional efficiency $\epsilon_{\text{pair}}=0.5$ for reconstructing and identifying the conversion pair as a phenomenological benchmark, motivated by converted-photon reconstruction at BABAR~\cite{BaBar:2011xka}.
The Supplemental Material specifies the generation settings, material and conversion weights, and detector-response parameters.

We reconstruct a candidate DV for each of the two possible recoil-photon assignments to the neutral ECL clusters ($\gamma_{\text{r}}$ and $\gamma_{\text{unc}}$).
For each assignment, the inferred ALP four-momentum, $p_a=P_{\text{ini}}-k_{\gamma_{\text{r}}}$, defines a flight line from the IP.
The measured conversion vertex and converted-photon direction determine the photon trajectory.
We reconstruct the DV as the midpoint of the shortest segment connecting these lines in three dimensions.
The segment's length is their distance of closest approach (DCA), $d_{\text{DCA}}^{a\gamma}$.
We select the assignment with the smaller chi-square-like score
\begin{equation}
 \chi^2_{\text{assn}}  =\chi^2_{\text{line}}+\chi^2_{4p}+\chi^2_{\text{recoil}}.
 \label{eqn:assignment_score}
\end{equation}
The three terms measure line compatibility, four-momentum closure, and consistency of the recoil mass squared $(P_{\text{ini}}-k_{\gamma_{\text{r}}})^2$ with the tested value $m_{\text{hyp}}^2\equiv m_a^2$, respectively.
We require $\chi^2_{\text{assn}} \le 25$ and $d_{\text{DCA}}^{a\gamma} \le 2.0$~mm.
The Supplemental Material defines the score and compares the reconstruction with MC truth.

At $(m_a,c\tau_a)=(100~\text{MeV},2~\text{mm})$, the full-VXD sample after energy and assignment cuts has an assignment purity of $99.74\%$.
In 68\% of these events, the reconstructed DV lies within 26.4~mm of the true DV in three dimensions.
The corresponding diphoton-mass resolution is 13.6~MeV.

We require reconstructed lab-frame energies of $E_{\gamma_\text{r}} \ge 1.0$~GeV for the recoil photon and $E_{\gamma_{\text{c}}},E_{\gamma_{\text{unc}}} \ge 0.1$~GeV for the daughters.
The chosen solution must be causal and lie within $20~{\text{mm}}\le R_{xy}^{\text{DV}}<1250$~mm and $-1020~{\text{mm}}<z^{\text{DV}}<1960$~mm, where $R_{xy}^{\text{DV}}$ and $z^{\text{DV}}$ are the DV transverse radius and longitudinal coordinate.
We also require the reconstructed diphoton invariant mass $m_{\gamma\gamma}$ to satisfy $|m_{\gamma\gamma}-m_{\text{hyp}}| \le 30$~MeV around the tested ALP mass $m_{\text{hyp}}$.
We suppress prompt backgrounds by requiring the converted-photon trajectory to be displaced from the IP.
We quantify this displacement by the line's transverse DCA to the IP, $d_{\text{DCA}}^{\gamma}$, and require $d_{\text{DCA}}^{\gamma} \ge 1.5$~mm for all mass and lifetime hypotheses.
This requirement suppresses the prompt $3\gamma$ background by nearly four orders of magnitude relative to the pre-DCA selection.
The Supplemental Material provides a reconstruction schematic, analytic closest-approach expressions, and representative $d_{\text{DCA}}^{\gamma}$ threshold scans.

\begin{figure}[t]
    \centering
    \includegraphics[width=0.99\linewidth]{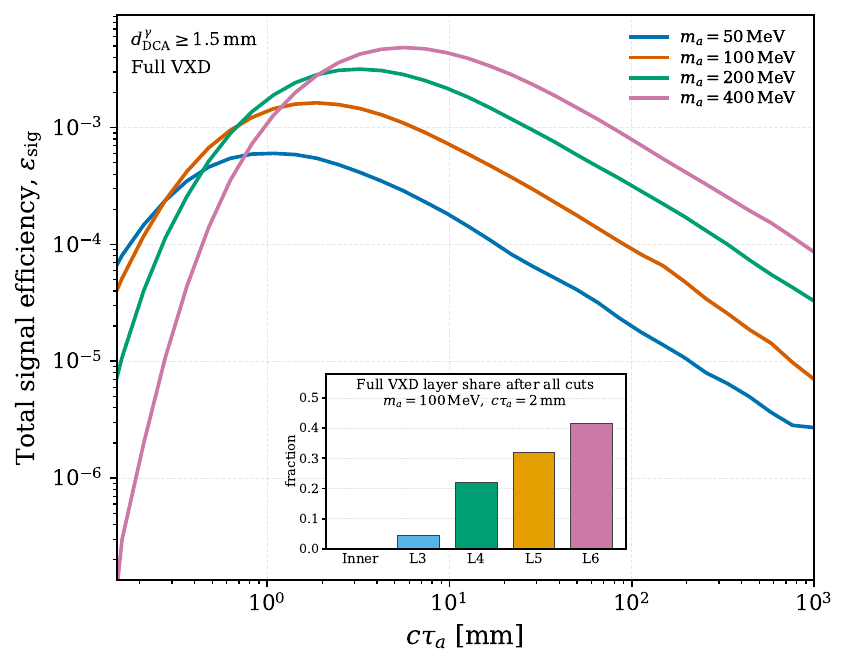}
    \caption{Total signal efficiency after the complete event selection versus $c\tau_a$ for the full VXD, including the $d_{\text{DCA}}^{\gamma}\ge1.5$~mm requirement, conversion and survival probabilities, and $\epsilon_{\text{pair}}=0.5$. The inset shows the selected conversion-layer composition at $m_a=100$~MeV and $c\tau_a=2$~mm; ``Inner'' combines the beam pipe and the two PXD layers.}
    \label{fig:efficiency_vs_ctau}
\end{figure}
The total signal efficiencies in Fig.~\ref{fig:efficiency_vs_ctau} peak when many ALPs decay before the outer VXD layers with a resolvable displacement from the IP.
The two detector configurations have nearly identical efficiencies because conversions in PXD~L2 contribute little after all selections.
At $m_a=100$~MeV and $c\tau_a=2$~mm, conversions in the outer SVD layers dominate the selected sample, as shown in the inset.

Prompt QED production, $e^+e^- \to 3\gamma$, is the dominant simulated background.
After one photon converts, its reconstructed topology closely resembles the signal: one conversion pair and two neutral photon candidates, with no additional hard final-state activity.
Because all three photons are produced promptly at the IP, a correctly reconstructed converted-photon trajectory should point back to the IP.
Such events enter the displaced signal region only when detector resolution generates an apparent photon displacement, while the reconstructed event also satisfies the recoil and kinematic consistency requirements.
The requirement $d^\gamma_{\text{DCA}} \ge 1.5$~mm provides the strongest single suppression of this background, reducing the prompt-$3\gamma$ rate by nearly four orders of magnitude relative to the complete pre-DCA selection.
The recoil assignment, four-momentum closure, and recoil-mass consistency provide additional rejection of kinematically incompatible configurations.
We use the complete tree-level matrix elements for the multiphoton backgrounds, without the additional electron/positron ISR structure functions or \texttt{Pythia} shower used for the signal.

\begin{figure}[t]
    \centering
    \includegraphics[width=0.99\linewidth]{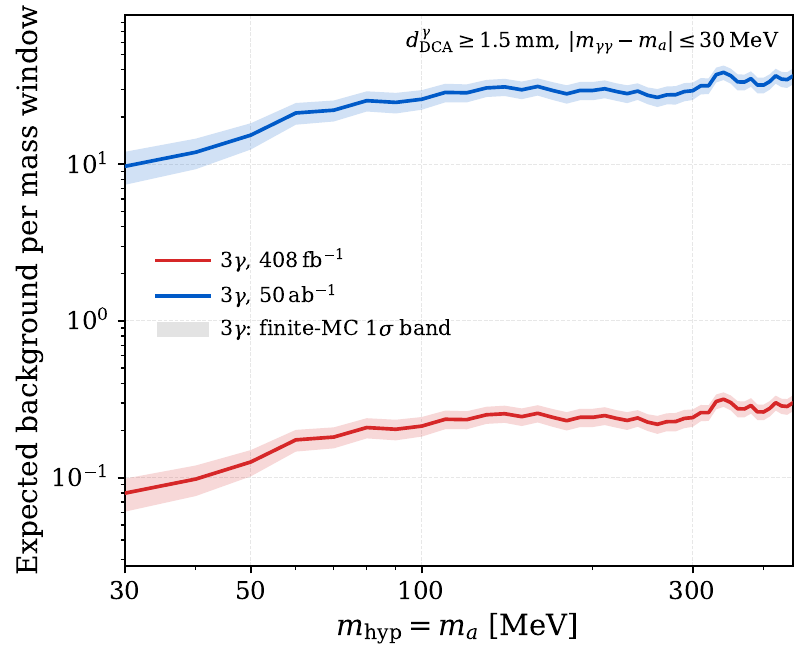}
    \caption{Expected prompt-$3\gamma$ background level in each $|m_{\gamma\gamma}-m_{\text{hyp}}| \le 30$~MeV window after all selections, for $408$~fb$^{-1}$ with the Run~1 VXD and $50$~ab$^{-1}$ with the full VXD. Shaded bands show the finite-MC $1\sigma$ uncertainty. The $4\gamma$, $\pi^0\gamma$, and reducible backgrounds are not included.}
    \label{fig:NumberOfBackgroundEvents_vs_ma}
\end{figure}

After all selections, the expected $3\gamma$ yield ranges from approximately $0.09$ to $0.28$ events per mass hypothesis for $408$~fb$^{-1}$ and from $10.6$ to $33.9$ events for $50$~ab$^{-1}$, as shown in Fig.~\ref{fig:NumberOfBackgroundEvents_vs_ma}.
The broad mass dependence reflects the reconstructed diphoton-mass spectrum of the three-photon background after recoil assignment and event selection.

The other simulated backgrounds are less important because of their smaller production rates and poorer compatibility with the reconstructed three-photon signal topology.
For $e^+e^-\to4\gamma$, an omitted fourth photon contributes to the four-momentum imbalance, and the selected yield is consequently only a few percent of the $3\gamma$ contribution.
The resonant channel $e^+e^-\to\pi^0\gamma$, $\pi^0\to\gamma\gamma$, produces a genuine diphoton resonance near $m_{\pi^0}$, but its much smaller production rate and prompt origin strongly suppress it after the displaced selection.
Its selected yield is below $3\times10^{-4}$ of the selected $3\gamma$ yield.
The Supplemental Material gives the generation details and normalization tests.

We do not simulate reducible backgrounds such as radiative Bhabha scattering and accidental or beam-induced combinations, whose residual rates depend on detector-specific reconstruction and beam conditions.
Experimentally, they can be suppressed using conversion-pair and material-vertex requirements, together with track and calorimeter vetoes and event-level consistency checks.
The Supplemental Material discusses these rejection handles in more detail and quantifies the possible impact of unmodeled backgrounds by increasing the nominal background rate by factors of 10 and 100.

\vspace{0.2cm}
\emph{\textbf{Sensitivity reach}---}
For each ALP mass and lifetime hypothesis, the expected signal yield is
\begin{equation}
N_{\text{sig}}=\mathcal{L}_{\text{int}}\cdot\sigma(e^+e^-\to a\gamma)\cdot
\text{BR}(a\to\gamma\gamma)\cdot \epsilon_{\text{sig}}(m_a,c\tau_a),
\label{eqn:signal_yield}
\end{equation}
where the ALP proper decay length $c\tau_a=\hbar c/\Gamma_a\simeq\hbar c/\Gamma(a\to\gamma\gamma)$ depends on $m_a$ and $g_{a\gamma\gamma}$, with $\Gamma_a$ denoting the total ALP decay width.

We treat the event count in each diphoton-mass window as a single-bin counting experiment.
We derive the expected $95\%$ CL exclusion using a one-sided profile-likelihood $\text{CL}_s$ construction under the background-only hypothesis~\cite{Read:2002hq,Cowan:2010js}.
A constrained nuisance parameter accounts for the finite-MC uncertainty in the prompt-$3\gamma$ background normalization.
The Supplemental Material details the likelihood and contour interpolation.

\begin{figure}[t]
\centering
\includegraphics[width=0.99\linewidth]{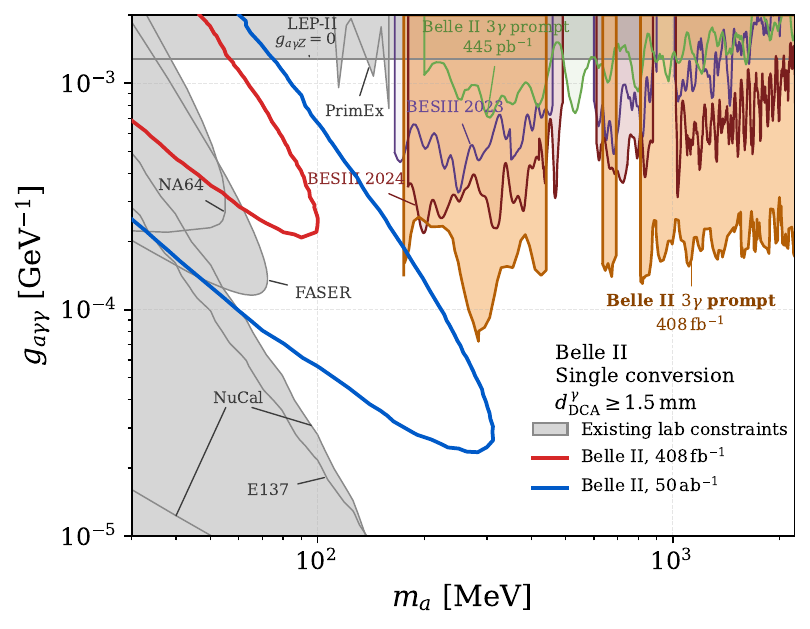}
\caption{Expected $95\%$ CL exclusion contours for $408$~fb$^{-1}$ with the Run~1 VXD configuration (red) and $50$~ab$^{-1}$ with the full VXD configuration (blue). Shaded regions denote existing laboratory exclusions~\cite{Knapen:2016moh,Dolan:2017osp,Aloni:2019ruo,NA64:2020qwq,Belle-II:2020jti,BESIII:2022rzz,BESIII:2024hdv,FASER:2024bbl,Belle-II:2026znr}.}
\label{fig:exclusion}
\end{figure}

The sensitivity contours for $408$~fb$^{-1}$ and $50$~ab$^{-1}$ in Fig.~\ref{fig:exclusion} close because both production and decay limit the selected signal yield.
At small $g_{a\gamma\gamma}$, both the production cross section and the probability for the ALP to decay within the detector decrease.
At large $g_{a\gamma\gamma}$, the ALP decays too promptly to satisfy the displaced-vertex requirements.

The recoil-assisted DV search probes the intermediate-lifetime gap between prompt collider searches and long-baseline fixed-target or forward-detector searches.
The existing $408$~fb$^{-1}$ Belle~II data set could exclude previously unconstrained parameter space around $(m_a,g_{a\gamma\gamma})\sim(100~\text{MeV},2\times10^{-4}~\text{GeV}^{-1})$, exceeding the existing bounds by nearly one order of magnitude.
With $50~\text{ab}^{-1}$, the nominal projection reaches couplings down to $g_{a\gamma\gamma} \simeq 2 \times 10^{-5}~\text{GeV}^{-1}$ and masses up to about 310~MeV.
The high-luminosity projection also covers the $\pi^0$-resonance window vetoed in the Belle~II prompt search~\cite{Belle-II:2026znr}.
The displaced-vertex requirements recover sensitivity in this window by suppressing the peaking $\pi^0\to\gamma\gamma$ background.

We test the detector and background assumptions by varying the parameterized model with all cuts fixed.
The projected reach is most sensitive to the converted-photon angular resolution.
We also find that even a hundredfold increase of the nominal background leaves some sensitivity beyond the existing laboratory constraints.
The Supplemental Material gives the individual variations and quantitative results.

\vspace{0.2cm}
\emph{\textbf{Conclusions and outlook}---}
We have introduced a recoil-assisted single-conversion strategy for three-dimensional vertexing of displaced diphoton decays at electron-positron colliders.
The known initial state and measured recoil-photon momentum define the LLP flight line.
One converted daughter then supplies the remaining directional information needed to locate the DV.
The conversion-related efficiency therefore scales linearly rather than quadratically with the photon-conversion probability.

For photophilic ALPs at Belle~II, our parameterized study finds that the existing $408$~fb$^{-1}$ data set could probe previously unconstrained parameter space near $m_a \simeq 100$~MeV and $g_{a\gamma\gamma} \sim 2\times10^{-4}$~GeV$^{-1}$.
With $50$~ab$^{-1}$, the reach extends to $m_a \simeq 310$~MeV and $g_{a\gamma\gamma} \simeq 2\times 10^{-5}$~GeV$^{-1}$.
These projections target the intermediate-lifetime region between prompt collider and displaced long-baseline searches.
The nominal high-luminosity projection also recovers sensitivity in the $\pi^0$-resonance window vetoed by the Belle~II prompt search~\cite{Belle-II:2026znr}.
Displaced vertexing can thus probe mass windows obscured by prompt Standard Model resonances.

The method can in principle extend to other electron-positron colliders such as BESIII and the proposed STCF and FCC-ee, provided they have a known initial state and suitable photon and conversion reconstruction.
It also applies to other neutral LLPs produced with a photon and decaying into photons.

The detector capabilities determine how to measure the daughter-photon direction.
At Belle~II, the homogeneous CsI(Tl) ECL does not independently measure the photon shower axis~\cite{Belle-II:2010dht}, so the converted daughter supplies the required three-dimensional pointing information.
At detectors with precise three-dimensional calorimeter pointing, a trajectory inferred from the calorimeter could replace the conversion measurement.
Measuring both daughter trajectories would remove the recoil requirement and permit an analogous fully calorimetric strategy at hadron colliders~\cite{ATLAS:2023meo}.

A quantitative experimental validation, beyond the scope of this phenomenological study, will ultimately require full detector simulation, measured conversion efficiencies, and data-driven estimates of reducible backgrounds.

\vspace{0.2cm}
\emph{\textbf{Acknowledgments}---}
We thank Xiaorong Zhou for useful discussions.
We used GPT-6 Astra (OpenAI) under our supervision to assist with aspects of the statistical methodology.
We independently reviewed and verified all AI-assisted outputs incorporated into this work.
This work was supported by the National Natural Science Foundation of China under Grants No.~12475106 and No.~12505120, and the Fundamental Research Funds for the Central Universities under Grant No.~JZ2025HGTG0252.

\bibliography{refs}

\clearpage
\onecolumngrid

\makeatletter
\onecolumn@grid@setup
\let\set@footnotewidth\set@footnotewidth@one
\let\compose@footnotes\compose@footnotes@one
\makeatother
\begin{center}
   \textbf{\large SUPPLEMENTAL MATERIAL \\[.2cm] ``Vertexing displaced diphoton decays with recoil photons at Belle II''}\\[.2cm]
  \vspace{0.05in}
  {Zeren Simon Wang and Yu Zhang}
\end{center}
\setcounter{equation}{0}
\setcounter{figure}{0}
\setcounter{table}{0}
\setcounter{page}{1}
\setcounter{section}{0}
\setcounter{secnumdepth}{1}
\makeatletter
\renewcommand{\thesection}{S-\Roman{section}}
\renewcommand{\theequation}{S-\arabic{equation}}
\renewcommand{\thefigure}{S-\arabic{figure}}
\renewcommand{\thetable}{S-\arabic{table}}
\renewcommand{\bibnumfmt}[1]{[S#1]}
\renewcommand{\citenumfont}[1]{#1}
\renewcommand{\theHequation}{S.\arabic{equation}}
\renewcommand{\theHfigure}{S.\arabic{figure}}
\renewcommand{\theHtable}{S.\arabic{table}}
\renewcommand{\theHsection}{S.\arabic{section}}
\makeatother

\section{Simulation framework and detector model}
\label{sec:supp_simulation}

\subsection{Signal kinematics, generation, and lifetime treatment}
For $e^+e^-\to a \gamma$ with unpolarized beams and negligible electron mass, the ALP kinematics in the COM frame are
\begin{equation}
\frac{1}{\sigma}\frac{d\sigma}{d\cos{\theta_a^*}}=\frac{3}{8}(1+\cos^2{\theta_a^*}),  \qquad E_a^*=\frac{s+m_a^2}{2\sqrt{s}},  \quad p_a^*\equiv |\boldsymbol{p}_a^*|=\frac{s-m_a^2}{2\sqrt{s}}.
\label{eqn:differential_production_XS}
\end{equation}
For head-on beams\footnote{We neglect the SuperKEKB beam-crossing angle in these analytic expressions.} with lab-frame energies $E_-$ and $E_+$ of the electrons and positrons ($E_- > E_+$), respectively, the COM-to-lab boost has speed $\beta_B=(E_--E_+)/(E_-+E_+)$ and Lorentz factor $\gamma_B=(E_-+E_+)/\sqrt{s}$.
Writing $z\equiv\cos\theta_a^*$, we obtain the lab-frame ALP three-momentum magnitude
\begin{equation}
 |\boldsymbol{p}_a^{\text{lab}}(z)|=\left[(p_a^*)^2(1-z^2)+\gamma_B^2(p_a^*z+\beta_BE_a^*)^2\right]^{1/2},
\end{equation}
and the mean decay length averaged over the COM-frame production angle,
\begin{equation}
\left\langle\lambda_a^{\text{lab}}\right\rangle = \frac{\hbar c}{m_a \Gamma_a} \int_{-1}^{1}dz\,\frac{3}{8}(1+z^2)|\boldsymbol{p}_a^{\text{lab}}(z)|\equiv \frac{\hbar c}{m_a \Gamma_a}\left\langle |\boldsymbol{p}_a^{\text{lab}}|\right\rangle.
\label{eqn:supp_decay_length_angle_average}
\end{equation}
For $m_a\ll\sqrt{s}$, the even angular distribution gives $\langle|\boldsymbol{p}_a^{\text{lab}}|\rangle\simeq\gamma_BE_a^* \simeq (E_-+E_+)/2=5.5$~GeV.
At $(m_a,g_{a\gamma\gamma})=(100~\text{MeV},10^{-4}~\text{GeV}^{-1})$, we have $c\tau_a=3.97$~mm and $\langle\lambda_a^{\text{lab}}\rangle\simeq218$~mm, reproducing Eq.~(5) in the main text.

We generate $e^+e^-\to a\gamma, a\to\gamma\gamma$ signal samples with \texttt{MadGraph5\_aMC\_v3.6.6}~\cite{Alwall:2014hca,Frederix:2018nkq} using a UFO model implemented in \texttt{FeynRules~2.3.49}~\cite{Christensen:2008py,Alloul:2013bka}.
We simulate $10^5$ events at each mass point from 30 to 450~MeV in 10-MeV steps, without generator-level cuts.
We fix $g_{a\gamma\gamma}$ at $1$~GeV$^{-1}$ during generation and rescale the cross section reported in the LHE files to other couplings using its analytic $g_{a\gamma\gamma}^2$ dependence.
The narrow-width approximation is valid throughout this mass grid.

We include electron and positron ISR structure functions by setting \texttt{lpp1=+3}, \texttt{lpp2=-3}, and \texttt{pdlabel1=pdlabel2=isronlyll} in the signal generation.
We then pass the LHE events to \texttt{Pythia~8.3}~\cite{Sjostrand:2014zea,Bierlich:2022pfr} with the lepton-QED ISR shower enabled.
We disable QCD showering, final-state radiation, multiparton interactions, and hadronization.

We scan 60 logarithmically spaced lifetimes between $c\tau_a=10^{-4}$ and $10^3$~mm by rescaling the sampled decay positions.
For each event, $\ell_{\text{prop}}^{\text{LHE}}$ is the ALP proper decay time multiplied by $c$, as recorded in the LHE file.
For a sample generated with mean proper decay length $c\tau_a^{\text{gen}}$, we rescale this quantity to each target $c\tau_a$ as
\begin{equation}
\ell_{\text{prop}}(c\tau_a)=\ell_{\text{prop}}^{\text{LHE}}\frac{c\tau_a}{c\tau_a^{\text{gen}}},\qquad  \boldsymbol{x}_{\text{dec}}(c\tau_a)=\frac{\boldsymbol{p}_a^{\text{lab}}}{m_a}\ell_{\text{prop}}(c\tau_a).
\label{eqn:supp_lifetime_rescale}
\end{equation}
This rescaling preserves each event's dimensionless ratio $\ell_{\text{prop}}/(c\tau_a)$ without an additional lifetime-reweighting factor.
At each target lifetime, we keep the event kinematics fixed and recalculate the material and ECL intersections and the conversion and survival weights from the rescaled decay position.
We then repeat the detector response, reconstruction, and selection, and we average the accepted weights over the original event sample.

\subsection{Geometry, material, and parameterized response}
The ECL accepts prompt photons within $12.4^\circ<\theta<31.4^\circ$, $32.2^\circ<\theta<128.7^\circ$, or $130.7^\circ<\theta<155.1^\circ$~\cite{Belle-II:2023cal}.
Following the Belle~II detector geometry~\cite{Belle-II:2010dht}, we model the ECL front surface by a cylindrical barrel of radius $R_{\text{ECL}}=1250$~mm spanning $-1020<z<1960$~mm and endcap planes at $z=1960$~mm and $z=-1020$~mm.
Combined with the angular acceptance above, these correspond to effective radial ranges $430<R_{xy}<1200$~mm and $470<R_{xy}<1190$~mm for the forward and backward endcaps, respectively.
For a displaced photon, we propagate a ray from its production point, i.e.~the ALP decay vertex, and require a positive-path intersection with one of these surfaces, taking the first intersection along the photon direction.

We represent the beam pipe, PXD, and SVD by the effective cylindrical shells in Table~\ref{table:detector_model_parameters}.
For a shell of radius $R_i$, we include a photon crossing only if its longitudinal coordinate satisfies $R_i/\tan150^\circ\le z_{\text{cross}}\le R_i/\tan17^\circ$.

\begin{table}[h]
\centering
\begin{tabular}{lcccc}
\hline\hline
Element & $R_i$ [mm] & $(x/X_0)_i$ & $f_i$ (Run~1) & $f_i$ (full VXD)\\
\hline
Beam pipe & 10  & $0.28\%$ & 1 & 1\\
PXD L1    & 14  & $0.20\%$ & 1 & 1\\
PXD L2    & 22  & $0.20\%$ & $1/6$ & 1\\
SVD L3    & 39  & $0.70\%$ & 1 & 1\\
SVD L4    & 80  & $0.70\%$ & 1 & 1\\
SVD L5    & 105 & $0.70\%$ & 1 & 1\\
SVD L6    & 135 & $0.70\%$ & 1 & 1\\
\hline\hline
\end{tabular}
\caption{Effective material-shell model based on the Belle~II detector geometry and reported layer-averaged material budgets~\cite{Belle-II:2010dht,Belle-IISVD:2022upf,Belle-IIDEPFET:2024iqz}.
$R_i$ is the shell radius, and $(x/X_0)_i$ is its assigned normal-incidence material budget.
We obtain the beam-pipe entry from the combined 1.0-mm thickness of its two beryllium walls and the beryllium radiation length in Ref.~\cite{Navas:2024}.
The PXD and SVD entries use the reported average budgets per layer, including sensor, support, and readout material.
The azimuthally averaged coverage factor $f_i$ equals unity for the beam pipe and a fully instrumented layer.
In Run~1, PXD~L2 contained two of its twelve nominal ladders, giving $f_i=1/6$.
We do not model individual ladder boundaries.}
\label{table:detector_model_parameters}
\end{table}

At an ordered shell crossing $i$, the path-length-corrected thickness, conversion probability, and first-conversion weight are
\begin{equation}
 t_i=s_{\text{mat}}\frac{f_i\cdot(x/X_0)_i}{|\hat{\boldsymbol{n}}_\gamma\cdot\hat{\boldsymbol{n}}_i|},\qquad    p_i=1-e^{-7t_i/9},\qquad w_i=p_i\prod_{j<i}(1-p_j).
\label{eqn:supp_layered_conversion_probability}
\end{equation}
Here $\hat{\boldsymbol{n}}_i$ is the radial shell normal at the crossing.
The factor $s_{\text{mat}}$ scales all assigned shell material budgets; we use $s_{\text{mat}}=1$ in the nominal model.
The factor $7/9$ follows from the high-energy pair-production mean free path $9X_0/7$~\cite{Navas:2024}.
The photon survival probability is $P_{\text{surv}}=\prod_i(1-p_i)$.
We evaluate each possible conversion separately.
For a retained three-photon system, a branch $(\text{c},\ell)$ specifies that photon $\text{c}$ first converts at material crossing $\ell$ while the other two photons remain unconverted.
Its probability is
\begin{equation}
P_{\text{c}\ell}^{\text{br}} = \epsilon_{\text{pair}}w_\ell^{(c)} \prod_{j\ne c}P_{\text{surv}}^{(j)}.
\label{eqn:supp_conversion_branch}
\end{equation}

For signal, only the two true ALP daughter photons are allowed to define the converted-photon branch.
All other final-state photons are required to remain unconverted.
Denoting by $\mathcal{X}$ any additional final-state photons beyond the recoil photon and the two ALP daughters, the summed single-conversion probability is
\begin{equation}
P_{\text{1conv}} = \epsilon_{\text{pair}} P_{\text{surv}}^{\text{r}} \prod_{k\in\mathcal{X}}P_{\text{surv}}^{(k)} \left[ P_{\text{surv}}^{(2)}\sum_i w_i^{(1)} + P_{\text{surv}}^{(1)}\sum_i w_i^{(2)} \right].
\label{eqn:supp_single_conversion_weight}
\end{equation}
Here the superscripts ``r'', ``$(1)$'', and ``$(2)$'' refer to the recoil photon and the two displaced photons, respectively.
For prompt-photon backgrounds, every \texttt{Pythia}-level photon is considered as a possible converting candidate; a branch contributes only if the converted candidate belongs to the reconstructed top three.

For each branch, we generate $N_{\text{rep}}$ detector replicas by sampling the parameterized response with the truth-level event and conversion configuration fixed.
We use $N_{\text{rep}}=5$ for signal, 300 for $3\gamma$, and 100 for $4\gamma$ and $\pi^0\gamma$.
Each replica carries weight $P_{\text{c}\ell}^{\text{br}}/N_{\text{rep}}$.
We repeat reconstruction and selection for each replica, choosing the recoil assignment from the two unconverted photons.

We model the detector response with Gaussian smearing of photon energies, neutral-cluster positions at the ECL, the converted-photon direction, and the conversion vertex.
Table~\ref{tab:supp_response} lists the nominal parameters.
Below, we explain their experimental motivation and limitations.

\begin{table}[H]
\centering
\begin{tabular}{lc}
\hline\hline
Item & Nominal resolution or assumption\\
\hline
Neutral ECL energy & $\sigma_E/E=\Big((0.02/\sqrt{E/\text{GeV}})^2+0.01^2\Big)^{1/2}$\\
Neutral ECL position & 10~mm in each local surface coordinate\\
Converted-photon energy & $\sigma_E/E=1\%$\\
Converted-photon direction & 1~mrad in each transverse angular component\\
Conversion vertex & $(\sigma_x,\sigma_y,\sigma_z)=(0.10,0.10,0.50)$~mm\\
Conditional conversion-reconstruction efficiency & $\epsilon_{\text{pair}}=0.5$\\
\hline\hline
\end{tabular}
\caption{Nominal parameters of the phenomenological detector-response model. The quoted resolutions are Gaussian standard deviations.}
\label{tab:supp_response}
\end{table}

We use published Belle ECL performance as a reference for the nominal smearing widths.
The reported energy resolutions are about $4\%$ at 100~MeV and $1.8\%$ at 1~GeV, with a position-resolution scale of $5~\text{mm}/\sqrt{E/\text{GeV}}$~\cite{Matvienko:2017ECL}.
Our energy widths are $6.4\%$ and $2.24\%$ at these energies.
The fixed 10-mm position width lies between the corresponding reference values of approximately 16 and 5~mm.

We assume unit reconstruction efficiency for neutral ECL clusters within the geometrical acceptance and apply the photon-energy requirements after smearing.
We do not model the incidence-angle dependence of the cluster response to non-pointing photons.

For converted photons, we smear the momentum and conversion vertex directly at photon level.
We take a constant relative energy width of $1\%$, guided by the improved photon-energy resolution obtained with conversion tracking at BaBar~\cite{BaBar:2011xka}.
We use a Gaussian width of 1~mrad in each transverse angular component as a conservative phenomenological benchmark for the converted-photon direction.
This choice is motivated by the Belle~II SVD's intrinsic hit resolution of ${\cal O}(10~\mu\text{m})$~\cite{Belle-IISVD:2022upf} and recent studies of resolving milliradian-scale conversion-pair opening angles at Belle~II~\cite{deLima:2026jbb}.

For the conversion vertex, we adopt the phenomenological widths $(\sigma_x,\sigma_y,\sigma_z)=(0.10,0.10,0.50)$~mm to allow for degradation from low-mass conversion-pair reconstruction and track extrapolation.
We test their impact by doubling all three widths in Sec.~\ref{sec:supp_robustness}.

We assume a constant conditional probability $\epsilon_{\text{pair}}=0.5$ for reconstructing and identifying a conversion pair, separate from the material conversion probability.
Section~\ref{sec:supp_robustness} examines how the converted-photon angular and vertex widths and $\epsilon_{\text{pair}}$ affect the projected sensitivity.

\section{Reconstruction and event selection}
\label{sec:supp_reconstruction}

\subsection{Vertex reconstruction and recoil assignment}

We reconstruct a candidate DV for each of the two recoil assignments and select the one with the smaller score.
Each single-conversion branch contains one converted photon and two neutral ECL clusters.
The converted-photon trajectory and its displacement from the IP are common to both assignments.

We work in the laboratory frame, with the IP at the origin and the $z$-axis along the beam, neglecting beam-spot smearing.
Let $\boldsymbol{r}_{\text{conv}}=(x_{\text{conv}},y_{\text{conv}},z_{\text{conv}})$ denote the reconstructed conversion vertex.
The unit vector $\hat{\boldsymbol{n}}_\gamma=(n_{\gamma x},n_{\gamma y},n_{\gamma z})$ gives the reconstructed forward direction of the converted photon.
All conversion-vertex and photon observables include the detector smearing described above.
The back-extrapolated converted-photon line is
\begin{equation}
 \boldsymbol{r}_\gamma(s)=\boldsymbol{r}_{\text{conv}}-s\hat{\boldsymbol{n}}_\gamma.
\end{equation}
Here $s$ is a signed distance parameter: positive $s$ extrapolates backward from the conversion vertex.

To define the photon displacement from the IP, we project $\boldsymbol{r}_{\text{conv}}$ and $\hat{\boldsymbol{n}}_{\gamma}$ onto the transverse $(x,y)$ plane:
\begin{equation}
\boldsymbol{r}_{\text{conv},\text{T}} = (x_{\text{conv}},y_{\text{conv}},0),\qquad  \hat{\boldsymbol{n}}_{\gamma,\text{T}}  =(n_{\gamma x},n_{\gamma y},0).
\end{equation}
The transverse distance of closest approach of the converted-photon line to the IP is
\begin{equation}
d_{\text{DCA}}^{\gamma}=  \frac{|\boldsymbol{r}_{\text{conv},\text{T}}  \times\hat{\boldsymbol{n}}_{\gamma,\text{T}}|}  {|\hat{\boldsymbol{n}}_{\gamma,\text{T}}|}.
\label{eqn:supp_photon_dca}
\end{equation}

We keep the line $\boldsymbol{r}_\gamma(s)$ and displacement $d_{\text{DCA}}^\gamma$ fixed while testing each neutral ECL cluster as the recoil photon, giving two hypotheses $h=1,2$.
In each hypothesis, we assign the other cluster as the unconverted-daughter candidate and evaluate the reconstruction and score separately.
We suppress the hypothesis label $h$ below.

We write the known initial-state four-momentum as $P_{\text{ini}}=(E_{\text{ini}},\boldsymbol{p}_{\text{ini}})$ and the reconstructed photon four-momenta as $k_i=(E_i,E_i\hat{\boldsymbol{n}}_i)$.
The index $i=\text{r},\text{c},\text{unc}$ labels the photons assigned as the recoil photon, converted daughter, and unconverted daughter, respectively.
Here $E_i$ is the smeared laboratory-frame energy and $\hat{\boldsymbol{n}}_i$ is the photon direction for that hypothesis; for the converted photon, $\hat{\boldsymbol{n}}_{\text{c}}=\hat{\boldsymbol{n}}_\gamma$.

The recoil direction points from the IP to the candidate cluster's smeared ECL position $\boldsymbol{x}_{\text{r}}^{\text{ECL}}$.
Its four-momentum determines the candidate ALP four-momentum:
\begin{equation}
\hat{\boldsymbol{n}}_{\text{r}}=  \frac{\boldsymbol{x}_{\text{r}}^{\text{ECL}}}  {|\boldsymbol{x}_{\text{r}}^{\text{ECL}}|},  \qquad p_a=P_{\text{ini}}-k_{\text{r}}.
\end{equation}
The three-momentum $\boldsymbol{p}_a$ defines the ALP flight direction $\hat{\boldsymbol{n}}_a=\boldsymbol{p}_a/|\boldsymbol{p}_a|$ and the line
\begin{equation}
 \boldsymbol{r}_a(L)=L\hat{\boldsymbol{n}}_a.
\end{equation}
Here $L$ is a signed distance parameter: positive $L$ follows the candidate ALP direction from the IP.

We define $c\equiv\hat{\boldsymbol{n}}_a\cdot\hat{\boldsymbol{n}}_\gamma$, $A\equiv\hat{\boldsymbol{n}}_a\cdot\boldsymbol{r}_{\text{conv}}$, and $B\equiv\hat{\boldsymbol{n}}_\gamma\cdot\boldsymbol{r}_{\text{conv}}$.
For nonparallel lines, minimizing $|\boldsymbol{r}_a(L)-\boldsymbol{r}_\gamma(s)|^2$ gives
\begin{equation}
L_{\text{sol}}=\frac{A-cB}{1-c^2},\qquad s_{\text{sol}}=\frac{B-cA}{1-c^2}.
\label{eqn:supp_closest_parameters}
\end{equation}
These expressions give the closest points on the unrestricted lines.
We use their separation as the three-dimensional line DCA and their midpoint as the reconstructed DV:
\begin{equation}
 d_{\text{DCA}}^{a\gamma}=|\boldsymbol{r}_a(L_{\text{sol}})-\boldsymbol{r}_\gamma(s_{\text{sol}})|,\qquad  \boldsymbol{r}_{\text{DV}}^{\text{reco}}= \frac{\boldsymbol{r}_a(L_{\text{sol}})+\boldsymbol{r}_\gamma(s_{\text{sol}})}{2}.
\label{eqn:supp_line_dca_dv}
\end{equation}

\begin{figure}[t]
\centering
\includegraphics[width=0.7\linewidth]{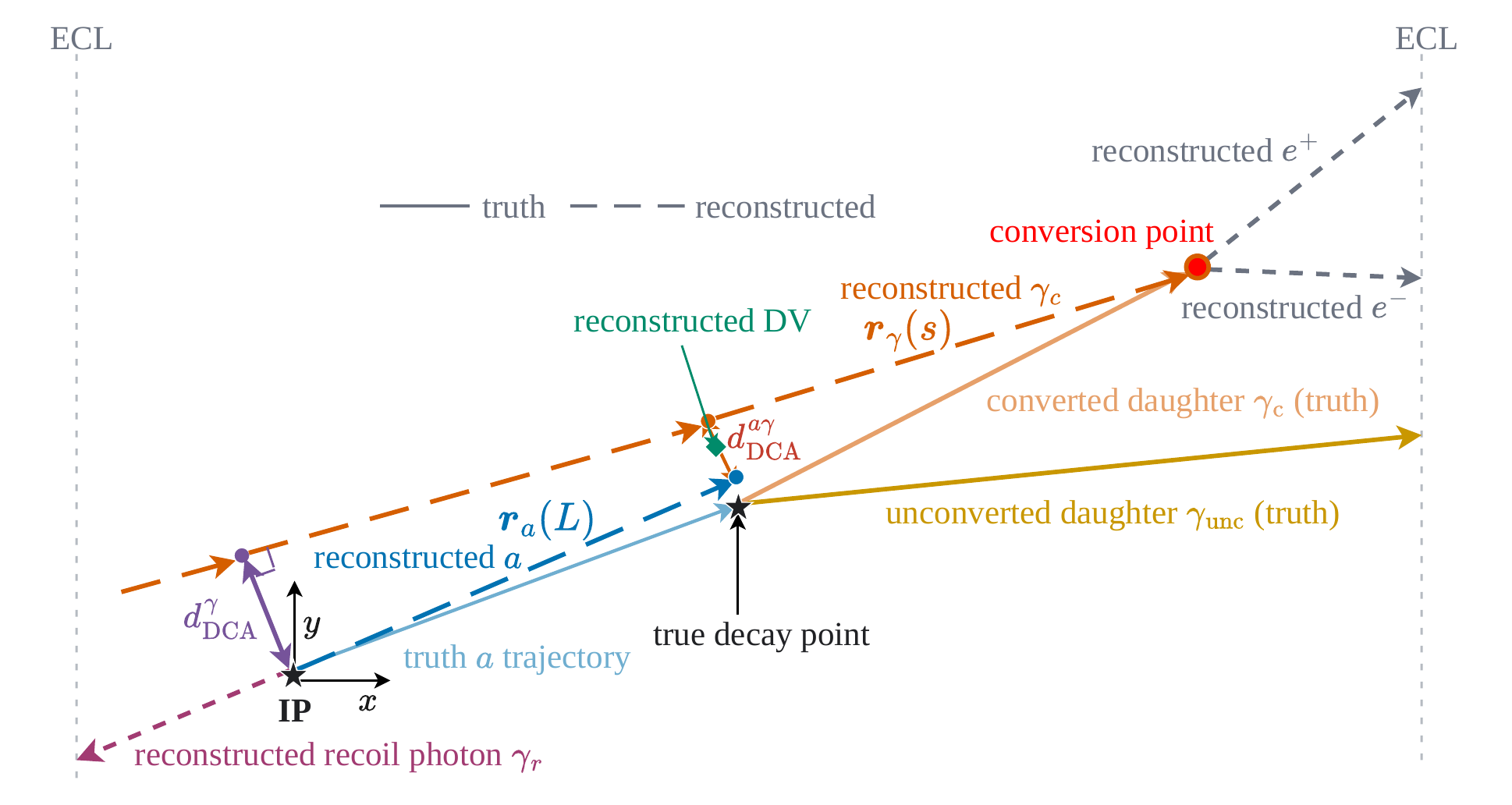}
\caption{Two-dimensional projection of the reconstruction. The DV estimator is the midpoint of the shortest three-dimensional segment between the ALP and converted-photon lines. The photon displacement $d_{\text{DCA}}^{\gamma}$ is the transverse distance of closest approach to the IP.}
\label{fig:supp_DCA}
\end{figure}
Fig.~\ref{fig:supp_DCA} illustrates the closest-approach construction and the two DCA definitions.

For each hypothesis, we determine the unconverted daughter's direction from the candidate DV to the remaining cluster's smeared ECL position $\boldsymbol{x}_{\text{unc}}^{\text{ECL}}$.
Together with the cluster energy, this gives the daughter's four-momentum and the diphoton mass squared:
\begin{equation}
 \hat{\boldsymbol{n}}_{\text{unc}}  =\frac{\boldsymbol{x}_{\text{unc}}^{\text{ECL}}  -\boldsymbol{r}_{\text{DV}}^{\text{reco}}}  {|\boldsymbol{x}_{\text{unc}}^{\text{ECL}}  -\boldsymbol{r}_{\text{DV}}^{\text{reco}}|},  \qquad  k_{\text{unc}}= \left(E_{\text{unc}},E_{\text{unc}}\hat{\boldsymbol{n}}_{\text{unc}}\right),  \qquad  m_{\gamma\gamma}^2=(k_{\text{c}}+k_{\text{unc}})^2.
\label{eqn:supp_unconverted_momentum}
\end{equation}

For each assignment, the four-momentum residual is
\begin{equation}
 \Delta P \equiv (\Delta E,\Delta\boldsymbol{p}) =P_{\text{ini}}-k_{\text{r}}-k_{\text{c}}-k_{\text{unc}},
\end{equation}
where $\Delta E\equiv E_{\text{ini}}-\sum_i E_i$ and $\Delta\boldsymbol{p}\equiv \boldsymbol{p}_{\text{ini}}-\sum_i E_i\hat{\boldsymbol{n}}_i$ are the laboratory-frame energy and three-momentum imbalances.
Both sums run over $i=\text{r},\text{c},\text{unc}$.
We define the recoil mass squared as $m_{\text{rec}}^2\equiv(P_{\text{ini}}-k_{\text{r}})^2$ and compare it with the tested squared ALP mass $m_{\text{hyp}}^2$.

We rank the assignments using the dimensionless score
\begin{equation}
\chi^2_{\text{assn}} \equiv \chi^2_{\text{line}} +\chi^2_{4p} +\chi^2_{\text{recoil}},
\label{eqn:supp_assignment_score}
\end{equation}
whose terms measure line compatibility, four-momentum closure, and recoil-mass consistency, respectively:
\begin{align}
\chi^2_{\text{line}} &\equiv \left( \frac{d_{\text{DCA}}^{a\gamma}}{\sigma_{\text{line}}} \right)^2,\\
\chi^2_{4p} &\equiv \frac{(\Delta E)^2+|\Delta\boldsymbol{p}|^2} {\sigma_{\text{kin}}^2},\\
\chi^2_{\text{recoil}} &\equiv \left[ \frac{m_{\text{rec}}^2-m_{\text{hyp}}^2}{\sigma_{m_{\text{rec}}^2}} \right]^2.
\label{eqn:supp_assignment_terms}
\end{align}

We calculate $\sigma_{\text{line}}$ for each event by propagating the reconstruction uncertainties to the separation of the ALP and converted-photon lines.
Let
\begin{equation}
\hat{\boldsymbol{u}} = \frac{ \hat{\boldsymbol{n}}_a\times\hat{\boldsymbol{n}}_\gamma}{|\hat{\boldsymbol{n}}_a\times\hat{\boldsymbol{n}}_\gamma|}
\end{equation}
denote the unit normal to the two reconstructed directions.
The conversion-vertex contribution projected onto this normal is
\begin{equation}
\sigma_{\text{vtx},u}^2 = \sigma_{\text{vtx},xy}^2 (u_x^2+u_y^2) + \sigma_{\text{vtx},z}^2 u_z^2.
\end{equation}
We obtain the ALP-direction uncertainty from $\boldsymbol{p}_a=\boldsymbol{p}_{\text{ini}} -E_{\text{r}}\hat{\boldsymbol{n}}_{\text{r}}$.
Propagating the ECL position uncertainty gives the recoil-direction variance $\sigma_{\hat n_{\text{r}},u}^2 \equiv \mathrm{Var}(\delta\hat{\boldsymbol{n}}_{\text{r}} \cdot\hat{\boldsymbol{u}})$ along the normal.
The ALP-direction variance along $\hat{\boldsymbol{u}}$ is then
\begin{equation}
\sigma_{\hat n_a,u}^2 = \frac{ \sigma_{E_{\text{r}}}^2 (\hat{\boldsymbol{u}}\cdot\hat{\boldsymbol{n}}_{\text{r}})^2 + E_{\text{r}}^2\sigma_{\hat n_{\text{r}},u}^2 }{ |\boldsymbol{p}_a|^2}.
\end{equation}
The resulting propagated line-distance variance is
\begin{equation}
\sigma_{\text{line}}^2 = \sigma_{\text{vtx},u}^2 + L_{\text{sol}}^2\sigma_{\hat n_a,u}^2 + s_{\text{sol}}^2\sigma_{\theta_{\text{c}}}^2.
\label{eqn:supp_sigma_line}
\end{equation}
This variance accounts for the event geometry: angular uncertainties contribute more strongly when the ALP flight or converted-photon extrapolation distance is longer.
We use the response parameters in Table~\ref{tab:supp_response}.

We estimate the common kinematic scale $\sigma_{\text{kin}}$, in GeV, as
\begin{equation}
\sigma_{\text{kin}}^2\simeq \sigma_{E_{\text{r}}}^2+ \sigma_{E_{\text{unc}}}^2+ \sigma_{E_{\text{c}}}^2+ (E_{\text{r}}\sigma_{\theta_{\text{r}}})^2+ (E_{\text{unc}}\sigma_{\theta_{\text{unc}}})^2+ (E_{\text{c}}\sigma_{\theta_{\text{c}}})^2.
\end{equation}
Here $\sigma_{E_i}$ is the absolute energy width from the detector-response model and $\sigma_{\theta_i}$ is the corresponding angular scale in radians.
For neutral clusters, we use
\begin{equation}
\sigma_{\theta_{\text{r}}} = \frac{\sigma_{\text{ECL}}^{\text{pos}}}{L_{\text{r}}}, \qquad \sigma_{\theta_{\text{unc}}} = \frac{\sigma_{\text{ECL}}^{\text{pos}}}{L_{\text{unc}}},
\end{equation}
where $L_{\text{r}}=|\boldsymbol{x}_{\text{r}}^{\text{ECL}}|$ is the distance from the IP to the recoil cluster and $L_{\text{unc}} =|\boldsymbol{x}_{\text{unc}}^{\text{ECL}} -\boldsymbol{r}_{\text{DV}}^{\text{reco}}|$
is the distance from the candidate DV to the unconverted daughter's cluster.
We take the ECL position width $\sigma_{\text{ECL}}^{\text{pos}}$, converted-photon angular width $\sigma_{\theta_{\text{c}}}$, and photon-energy widths from Table~\ref{tab:supp_response}.

The measured recoil-photon energy and direction determine the recoil mass squared:
\begin{equation}
m_{\text{rec}}^2 = P_{\text{ini}}^2 - 2E_{\text{r}}
\left( E_{\text{ini}} - \boldsymbol{p}_{\text{ini}}\cdot \hat{\boldsymbol{n}}_{\text{r}} \right).
\end{equation}
Holding the initial-state four-momentum fixed, we propagate the recoil-photon energy and direction uncertainties to obtain an effective resolution:
\begin{equation}
\sigma_{m_{\text{rec}}^2}^2 = \Big[ 2\left| E_{\text{ini}} - \boldsymbol{p}_{\text{ini}}\cdot \hat{\boldsymbol{n}}_{\text{r}} \right| \sigma_{E_{\text{r}}} \Big]^2 + \Big( 2E_{\text{r}}|\boldsymbol{p}_{\text{ini}}| \sigma_{\theta_{\text{r}}} \Big)^2.
\label{eqn:supp_sigma_mrec2}
\end{equation}
The first term gives the energy contribution to the uncertainty in $m_{\text{rec}}^2$; the second approximates the directional contribution.
At first order, the angular coefficient also contains $\sin\theta_{\text{r}}$, where $\theta_{\text{r}}$ is the angle between $\boldsymbol{p}_{\text{ini}}$ and $\hat{\boldsymbol{n}}_{\text{r}}$.
We conservatively replace this factor by unity.
We neglect correlations among the energy, directional, and vertex uncertainties entering the estimates above.

The full-VXD benchmark at $m_a=m_{\text{hyp}}=100$~MeV and $c\tau_a=2$~mm illustrates the characteristic resolution scales entering the three assignment-score terms.
For signal, they are $\langle\sigma_{\text{line}}\rangle\simeq0.45$~mm, $\sqrt{\langle\sigma_{\text{kin}}^2\rangle}\simeq0.098$~GeV, and $\sqrt{\langle\sigma_{m_{\text{rec}}^2}^2\rangle}\simeq1.48$~GeV$^2$.
For the dominant prompt $3\gamma$ background at the same mass hypothesis, the corresponding values are approximately $0.28$~mm, $0.099$~GeV, and $1.43$~GeV$^2$.
Holding the residual fixed, these characteristic scales give background-to-signal $\chi^2$ ratios of approximately $2.6$, $0.98$, and $1.07$ for the line, four-momentum, and recoil-mass terms, respectively.
As a result, only the line term penalizes prompt background much more strongly than signal.

The actual score contributions also depend on the residuals.
For truth-correct signal assignments at this benchmark, the median $(\chi^2_{\text{line}},\chi^2_{4p},\chi^2_{\text{recoil}})$ is approximately $(0.58,2.11,0.73)$, so all three terms are of order unity.
For prompt $3\gamma$ at the same mass hypothesis, the pre-assignment means are $(0.96,1.97,5.7\times10^2)$.
The large mean recoil term reflects a tail of grossly incompatible background hypotheses that this term strongly penalizes.

We first select the hypothesis with the smaller score among those with a well-defined reconstruction.
We then apply the physical-solution requirements, including $L_{\text{sol}}>0$ and $s_{\text{sol}}>0$, and all remaining selections to the chosen hypothesis.
If it fails, we reject the detector replica without switching assignments.

\subsection{Ordered selection and DCA working point}

\begin{table}[t]
\centering
\begin{tabular}{p{0.30\linewidth}p{0.52\linewidth}}
\hline\hline
Selection variable & Requirement\\
\hline
Photon topology & Form detector-reconstructable photon candidates and retain the three with the highest smeared lab-frame energies; the converted-photon
candidate must be among them\\
ECL acceptance & Valid recoil-photon intersection within the stated ECL angular ranges and valid unconverted-daughter ray from the DV to the ECL\\
Single conversion & Exactly one retained photon first converts in an accepted shell; the other two survive the modeled material\\
Photon energies & $E_\gamma^{\text{r}} \ge 1.0$~GeV; $E_\gamma^{\text{c}},E_\gamma^{\text{unc}} \ge 0.1$~GeV\\
Recoil assignment & Valid nonparallel reconstruction with $\chi^2_{\text{assn}} \le 25$\\
Causality \& path-length limits & $0<L_{\text{sol}} \le 3000$~mm and $0<s_{\text{sol}} \le 3000$~mm\\
DV fiducial volume & $20~{\text{mm}}\le R_{xy}^{\text{DV}}<1250$~mm and $-1020~{\text{mm}}<z^{\text{DV}}<1960$~mm\\
Line compatibility & $d_{\text{DCA}}^{a\gamma}\le2.0$~mm\\
Diphoton mass & $|m_{\gamma\gamma}-m_{\text{hyp}}|\le30$~MeV\\
Photon displacement & $d_{\text{DCA}}^{\gamma}\ge1.5$~mm\\
\hline\hline
\end{tabular}
\caption{Baseline single-conversion selection. Photon energies, DV positions, and masses are reconstructed from lab-frame observables. Material and geometrical acceptance precede the reconstruction-level cut flow.}
\label{tab:supp_selection}
\end{table}

The baseline selection in Table~\ref{tab:supp_selection} uses only the three highest-energy photon candidates that the detector reconstructs.
For each signal conversion branch, the converted ALP daughter must be among
these candidates, while no truth requirement is imposed on the other two
retained candidates.
An ISR photon can therefore enter the retained three-photon system and is
handled by the same recoil-assignment procedure.
The unconverted daughter must intersect one of the modeled ECL surfaces when propagated from the truth decay point; the converted daughter does not require an ECL intersection.
We weight conversion branches as described in Sec.~\ref{sec:supp_simulation}.

We reject undefined or parallel reconstruction hypotheses and nonpositive reconstructed daughter paths to the ECL.
The requirements $L_{\text{sol}}>0$ and $s_{\text{sol}}>0$ enforce forward ALP propagation and backward photon extrapolation.
The 3-m upper limits reject excessively long extrapolations.

\begin{figure}[H]
\centering
\includegraphics[width=0.99\linewidth]{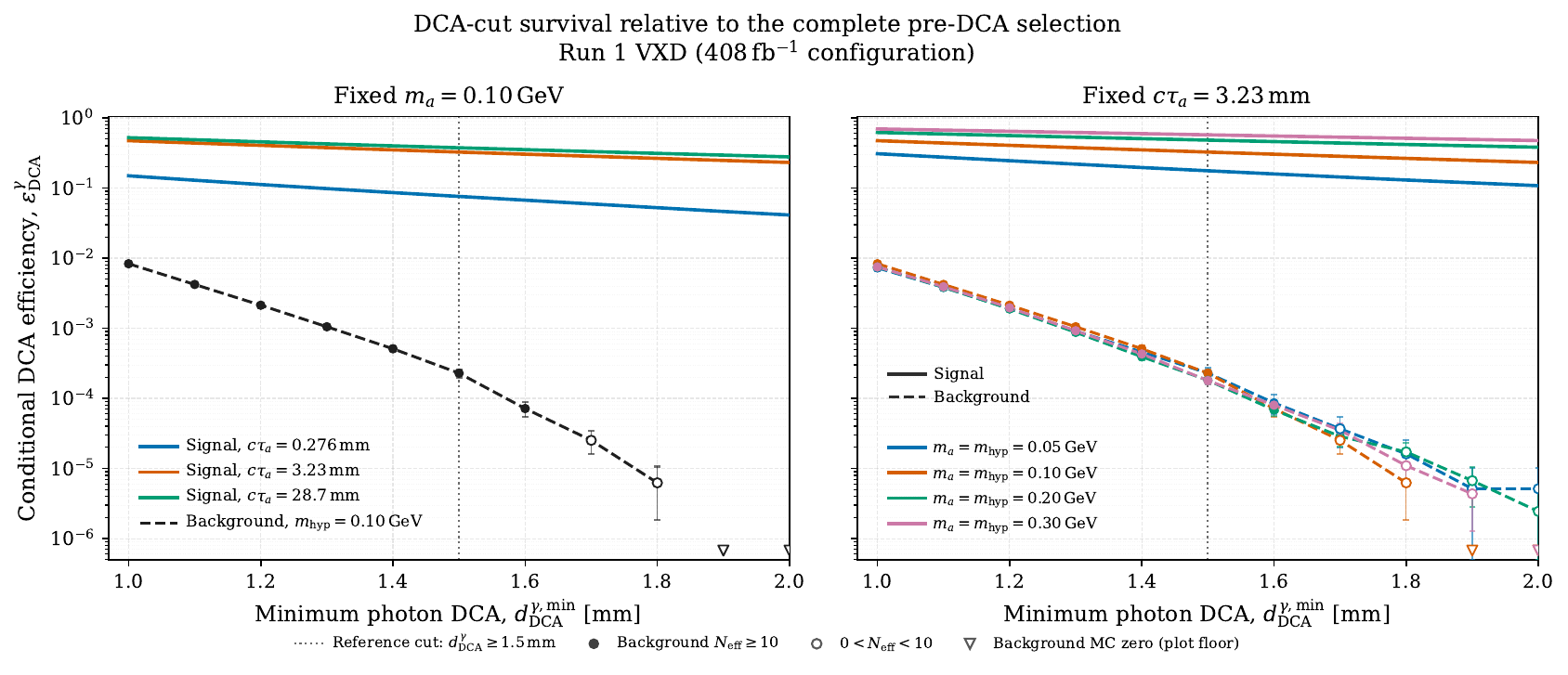}
\caption{Conditional survival probability of the $d_{\text{DCA}}^{\gamma}$ requirement for the Run~1 VXD.
Solid curves show signal and dashed curves with markers show prompt $3\gamma$.
The left panel fixes $m_a=0.10$~GeV and the right fixes $c\tau_a=3.23$~mm.
The dotted vertical line in each panel marks 1.5~mm.
Open markers flag low effective background statistics, and triangles mark scan points with no surviving background replica.}
\label{fig:supp_efficiency_dDCA}
\end{figure}

The common 1.5-mm DCA threshold suppresses the prompt $3\gamma$ background by nearly four orders of magnitude while retaining signal across the displayed benchmarks.
Fig.~\ref{fig:supp_efficiency_dDCA} shows the survival probabilities relative to the complete pre-DCA selection for the Run~1 VXD.
The full-VXD results are almost identical and are not shown.

\section{Signal reconstruction performance}
\label{sec:supp_signal_performance}

\subsection{Reconstruction validation}
Table~\ref{tab:supp_reco_validation} compares the full-VXD reconstruction with MC truth at $c\tau_a=2$~mm.
We evaluate these diagnostics after the energy and assignment requirements, before the DV, mass-window, and photon-DCA cuts.

\begin{table}[H]
\centering
\begin{tabular}{cccc}
\hline\hline
$m_a$ [MeV]~ & ~Assignment purity~ & ~DV resolution [mm]~ & ~$m_{\gamma\gamma}$ resolution [MeV]\\
\hline
50  & $99.73\%$ & 50.1 & 13.1\\
100 & $99.74\%$ & 26.4 & 13.6\\
200 & $99.74\%$ & 11.9 & 13.8\\
\hline\hline
\end{tabular}
\caption{Representative reconstruction performance for $c\tau_a=2$~mm. The DV resolution is defined as the three-dimensional distance from the true DV within which 68\% of the weighted reconstructed signal events lie. For the signed diphoton-mass residual $\Delta m_{\gamma\gamma}\equiv m_{\gamma\gamma}^{\text{reco}}-m_a$, $Q_p$ is the weighted $p^\text{th}$ percentile, and the mass resolution is the central-68\% half-width
$(Q_{84}-Q_{16})/2$.}
\label{tab:supp_reco_validation}
\end{table}

The DV resolution improves with increasing mass because the diphoton opening angle grows.
For all three benchmarks, the weighted median $Q_{50}$ of the diphoton-mass residual lies within 1.0~MeV of zero.

\subsection{Efficiency, cut flow, and MC precision}
We estimate each cumulative signal efficiency by averaging the accepted probability over all generated hard events.
At fixed $m_a$, $c\tau_a$, detector configuration, and selection stage, let $i=1,\ldots,N$ label these events.
For event $i$, the branch index $b\equiv(c,\ell)$ specifies the converting photon $c$ and its first-conversion material crossing $\ell$.
For event $i$, we evaluate the branch probability $P_{ib}^{\text{br}}$ using $P_{\text{c}\ell}^{\text{br}}$ from Eq.~\eqref{eqn:supp_conversion_branch}.
It includes the material-conversion probability, the other photons' survival probabilities, and $\epsilon_{\text{pair}}$.

For each branch, $r=1,\ldots,N_{\text{rep}}$ labels the detector-response replicas, with $N_{\text{rep}}=5$ for signal.
We set $I_{ibr}=1$ if replica $r$ of branch $b$ passes the baseline
requirements of Table~\ref{tab:supp_selection} through the chosen
selection stage, and $I_{ibr}=0$ otherwise.
The accepted-probability estimator for event $i$ is
\begin{equation}
 q_i\equiv
 \sum_b P_{ib}^{\text{br}}\,
 \frac{1}{N_{\text{rep}}}
 \sum_{r=1}^{N_{\text{rep}}}I_{ibr}.
\end{equation}
We assign $q_i=0$ to events with no allowed conversion branch or usable photon topology and retain them in the efficiency denominator.

Because the signal LHE samples are unweighted, the absolute cumulative efficiency and its finite-MC standard error are
\begin{equation}
\epsilon_{\text{sig}}=\frac{1}{N}\sum_{i=1}^{N}q_i,\qquad  \delta\epsilon_{\text{sig,MC}}=  \left[  \frac{1}{N(N-1)}  \sum_{i=1}^{N}(q_i-\epsilon_{\text{sig}})^2  \right]^{1/2}.
\label{eqn:supp_signal_efficiency_mc}
\end{equation}
We combine all branches and replicas into $q_i$ before evaluating the variance, so the statistically independent MC units are the generated hard events rather than the individual branches or replicas.

The relative MC uncertainty $\delta\epsilon_{\text{sig,MC}}/\epsilon_{\text{sig}}$ remains below approximately $10.5\%$ at grid points with $\epsilon_{\text{sig}} \ge 10^{-5}$.

\begin{table}[h]
\centering
\begin{tabular}{lccc}
\hline\hline
Selection stage & $c\tau_a=0.626$~mm & $3.23$~mm & $16.6$~mm\\
\hline
Material/topology and acceptance & $2.41\%$ & $1.24\%$ & $0.33\%$\\
$+\;$ photon energies & $2.31\%$ & $1.19\%$ & $0.32\%$\\
$+\;$ recoil assignment & $1.88\%$ & $0.94\%$ & $0.25\%$\\
$+\;$ causal DV and $d_{\text{DCA}}^{a\gamma}$ & $0.55\%$ & $0.47\%$ & $0.14\%$\\
$+\;$ diphoton mass & $0.52\%$ & $0.45\%$ & $0.13\%$\\
$+\;d_{\text{DCA}}^{\gamma}\ge1.5$~mm & $0.09\%$ & $0.15\%$ & $0.05\%$\\
\hline\hline
\end{tabular}
\caption{Absolute cumulative full-VXD efficiencies for $m_a=100$~MeV at three representative lifetime hypotheses, including $\epsilon_{\text{pair}}=0.5$. Each row includes all preceding requirements and is normalized to all generated signal events.}
\label{tab:supp_signal_cutflow}
\end{table}

The ``causal DV and $d_{\text{DCA}}^{a\gamma}$'' row combines the causality, path-length, DV fiducial-volume, and line-compatibility requirements in Table~\ref{tab:supp_selection}.
At the $(m_a,c\tau_a)=(100~\text{MeV},2~\text{mm})$ benchmark, approximately $95\%$ of the final selected signal weight arises from conversions in the three outer SVD layers.

\section{Background model and validation scope}
\label{sec:supp_background}

We use a fixed-order model for the multiphoton backgrounds.
We generate all background LHE samples without electron/positron ISR structure functions and analyze them with the \texttt{Pythia~8} lepton-QED ISR shower switched off.
We retain the complete tree-level $e^+e^- \to 3\gamma$ and $e^+e^- \to 4\gamma$ matrix elements, with all three or four photons in the hard final state.
Keeping the shower off avoids overlap with shower-generated photons in the absence of a matrix-element--shower matching prescription.
The nominal prediction does not include additional higher-order ISR effects on background rates or reconstructed distributions.

\subsection{\texorpdfstring{Prompt $3\gamma$ background}{Prompt three-photon background}}

After one photon converts, prompt $3\gamma$ events reproduce the same reconstructed three-photon topology as the signal, while detector resolution can generate an apparent displacement from the IP.
We generate five million unweighted $e^+e^- \to 3\gamma$ hard events with \texttt{MadGraph5\_aMC\_v3.6.6}, retaining the electron mass and setting $\alpha$ to $1/137$.
For each hard-process photon, we require $E_\gamma^*>50$~MeV in the COM frame and $|\eta_\gamma|<4.0$ in the lab frame, as configured in \texttt{MadGraph5}.
The resulting cross section is $\sigma_{3\gamma}=1.178$~nb.
We read the events through \texttt{Pythia~8} with all parton-level showering disabled.

Any of the three photons in a prompt $3\gamma$ event can convert.
For hard event $i$, a branch $b\equiv(c,\ell)$ specifies the converting photon $c$ and its first-conversion material crossing $\ell$.
We evaluate each branch with $N_{\text{rep}}=300$ detector-response replicas.
We set $I_{ibr}(m_{\text{hyp}})=1$ if replica $r$ of branch $b$ passes the complete selection at mass hypothesis $m_{\text{hyp}}$, and $I_{ibr}(m_{\text{hyp}})=0$ otherwise.
The accepted-probability estimator for event $i$ is
\begin{equation}
q_i(m_{\text{hyp}})  =  \sum_{b}  P_{ib}^{\text{br}}\,  \frac{1}{N_{\text{rep}}}  \sum_{r=1}^{N_{\text{rep}}}  I_{ibr}(m_{\text{hyp}}),
\end{equation}
where $P_{ib}^{\text{br}}$ is the branch probability defined in Eq.~\eqref{eqn:supp_conversion_branch}, evaluated for event $i$.
Events with no allowed branch have $q_i=0$.

The estimated selection efficiency and expected background yield are
\begin{equation}
 \widehat{\epsilon}_{3\gamma}(m_{\text{hyp}})  =  \frac{1}{N}\sum_{i=1}^{N}q_i(m_{\text{hyp}}),  \qquad  B_{3\gamma}(m_{\text{hyp}})  =  \mathcal{L}_{\text{int}} \cdot \sigma_{3\gamma} \cdot  \widehat{\epsilon}_{3\gamma}(m_{\text{hyp}}).
\end{equation}
We combine all branches and replicas of each hard event into $q_i$ before calculating the finite-MC uncertainty.
We treat each hard event as one statistically independent MC unit, giving the standard error
\begin{equation}
 \begin{aligned}
 \delta\epsilon_{3\gamma,\text{MC}}  &=  \Big(  \frac{1}{N(N-1)}  \sum_{i=1}^{N} \left(q_i-\widehat{\epsilon}_{3\gamma}\right)^2  \Big)^{1/2},\\
 \delta B_{3\gamma,\text{MC}}  &=  \mathcal{L}_{\text{int}} \cdot \sigma_{3\gamma} \cdot \delta\epsilon_{3\gamma,\text{MC}},  \qquad  N_{\text{eff},B}  =  \left(\frac{B_{3\gamma}}  {\delta B_{3\gamma,\text{MC}}}\right)^2.
 \end{aligned}
\label{eqn:supp_background_mc_error}
\end{equation}

As shown in Fig.~\ref{fig:supp_efficiency_dDCA}, the threshold $d^\gamma_{\text{DCA}} \ge 1.5$~mm reduces the prompt-$3\gamma$ background by nearly four orders of magnitude relative to the complete pre-DCA selection while retaining signal across the representative benchmarks.
After all selections, the expected background yield ranges from 0.09 to 0.28 events for $408$~fb$^{-1}$ and from 10.6 to 33.9 events for $50$~ab$^{-1}$ across the mass scan.
The corresponding relative finite-MC uncertainties range from $9.4\%$ to $18.6\%$.
Every mass hypothesis has a nonzero nominal background estimate.
The effective background statistics range from $N_{\text{eff},B}=28.9$ to 112.8, satisfying $N_{\text{eff},B} \ge 25$ throughout the scan.

\subsection{Subleading simulated backgrounds}
We generate one million tree-level $e^+e^- \to 4\gamma$ events with \texttt{MadGraph5\_aMC\_v3.6.6}, retaining the electron mass in the Standard Model matrix element.
We apply the same generator-level photon cuts as for $3\gamma$.
The resulting cross section for four resolved photons is $\sigma_{4\gamma}=90.81$~pb.
We read the events through \texttt{Pythia~8} with the same shower-off configuration.
We require at least four \texttt{Pythia}-level photons.
For each conversion branch, we form detector-reconstructable photon candidates and retain the three with the highest smeared lab-frame energies, following Table~\ref{tab:supp_selection}.
Any omitted photon is absent from the reconstructed three-photon system and therefore contributes to the four-momentum imbalance.
Within each detector replica, only the two retained neutral candidates contribute survival factors to the conversion-branch weight; no survival condition is imposed on an omitted photon.
We calculate the finite-MC uncertainty by the same event-level procedure as for $3\gamma$.

After selection, the resolved four-photon contribution is generally only a few percent of the selected $3\gamma$ yield, and we therefore omit it from the nominal background estimate.

For $e^+e^-\to\pi^0\gamma$, followed by $\pi^0\to\gamma\gamma$, we normalize the sample using
\begin{equation}
\sigma(e^+e^-\to\pi^0\gamma)  = \frac{2\pi^2\alpha^3(s)}{3}|F_{\pi\gamma}(s)|^2  \left(1-\frac{m_{\pi^0}^2}{s}\right)^3.
\end{equation}
As a normalization benchmark, we take $|sF_{\pi\gamma}(s)|\simeq\sqrt{2}f_\pi$, motivated by the spacelike asymptote $Q^2F_{\pi\gamma}(-Q^2)\to\sqrt{2}f_\pi$~\cite{Brodsky:1981rp}.
At $\sqrt{s}=10.58$~GeV, using $\alpha=1/137$, $f_\pi=130.2$~MeV, $m_{\pi^0}=134.98$~MeV, and $\text{BR}(\pi^0\to\gamma\gamma)=0.988$~\cite{Navas:2024}, we obtain $\sigma(e^+e^-\to \pi^0\gamma)\times\text{BR}(\pi^0\to\gamma\gamma) \simeq 2.7$~fb.

We generate one million events with a custom Python phase-space generator and no parton-level cuts.
We use the production-photon angular distribution $(3/8)(1+\cos^2\theta_\gamma^*)$ in the COM frame and an isotropic pion decay in its rest frame.
After the parameterized detector response and complete event selection in \texttt{Pythia~8}, the selected yield is below $3\times10^{-4}$ of the $3\gamma$ yield.
Even increasing the $\pi^0\gamma$ normalization by a factor of ten leaves this ratio below $0.3\%$, so uncertainty in the timelike form-factor normalization does not affect our conclusions.

\subsection{Reducible backgrounds}

We do not simulate radiative Bhabha scattering or accidental and beam-induced combinations.
An experimental analysis can suppress these backgrounds by requiring an opposite-sign, low-mass pair with a common vertex in detector material.
It can also veto additional tracks consistent with the IP and unexplained ECL activity.
Timing, conversion quality, and four-momentum closure provide further discrimination.
Estimating the residual rates requires full simulation with beam-background overlay and data control regions.
Section~\ref{sec:supp_robustness} quantifies the sensitivity to an assumed overall background increase.

\section{Statistical procedure}
\label{sec:supp_statistics}
For each mass hypothesis, we treat the events satisfying $|m_{\gamma\gamma}-m_{\text{hyp}}| \le 30$~MeV as a single-bin counting experiment.
Let $n$ be the count in this window, $s$ the tested signal yield from Eq.~(7) in the main text, and $B$ the nominal prompt-$3\gamma$ background yield.
We introduce a nonnegative background-normalization parameter $\beta$ to account for the finite-MC uncertainty $\delta B_{\text{MC}}$ in $B$:
\begin{equation}
 \mathcal{L}(s,\beta)=  \text{Pois}(n\mid s+\beta B)\,  \text{Pois}(n_{\text{aux}}\mid\beta N_{\text{eff},B}),  \qquad  N_{\text{eff},B}\equiv  \left(\frac{B}{\delta B_{\text{MC}}}\right)^2.
\label{eqn:supp_statistical_likelihood}
\end{equation}
At the nominal background-only point, we set the auxiliary observation to $n_{\text{aux}}=N_{\text{eff},B}$.
The auxiliary term therefore constrains $\beta$ around unity with relative width $1/\sqrt{N_{\text{eff},B}}=\delta B_{\text{MC}}/B$, following a Barlow--Beeston-type treatment of finite-MC uncertainty~\cite{Barlow:1993dm}.
For the weighted background sample, $N_{\text{eff},B}$ need not be an integer.
We therefore evaluate the auxiliary likelihood using the Gamma-function extension of the Poisson likelihood~\cite{Arguelles:2019izp}.

We obtain the expected upper limit $s_{95}$ at $95\%$ CL using the modified-frequentist $\text{CL}_s$ construction~\cite{Read:2002hq,Cowan:2010js}.
We use a one-sided profile-likelihood statistic for the nonnegative signal yield and profile over $\beta$.
We evaluate the expected limit with the background-only Asimov data, $(n,n_{\text{aux}})=(B,N_{\text{eff},B})$.
For mass windows with $B<10$, we calculate the sampling distributions entering $\text{CL}_s$ by discrete Poisson sums over both the signal-window and auxiliary counts.
For $B \ge 10$, we use the asymptotic profile-likelihood distribution~\cite{Cowan:2010js}.

At every simulated parameter point, we define
\begin{equation}
R_{\text{excl}}(m_a,g_{a\gamma\gamma}) = \frac{s(m_a,g_{a\gamma\gamma})}{s_{95}(m_a)}.
\end{equation}
We obtain the $95\%$ CL exclusion boundary by linearly interpolating $\log_{10}R_{\text{excl}}$ on the simulated grid in log mass and log coupling, without additional smoothing.

The background-normalization parameter $\beta$ is the only nuisance parameter in the nominal likelihood.

\section{Robustness of the projected sensitivity}
\label{sec:supp_robustness}
We test the detector and background assumptions by varying the parameterized model with all selections fixed.
For each detector variation, we recalculate the signal and prompt-$3\gamma$ background yields with the same hard-event samples, random-seed settings, and replica counts as in the nominal calculation.
We then recompute the limits using the procedure in Sec.~\ref{sec:supp_statistics}, including the corresponding finite-MC background constraint.

\subsection{Detector response, material, and conversion efficiency}
We vary one assumption at a time, starting from the nominal response in Table~\ref{tab:supp_response} and the material model in Table~\ref{table:detector_model_parameters} with $s_{\text{mat}}=1$.
In separate tests, we double the converted-photon angular width from 1 to 2~mrad, the conversion-vertex widths from $(0.10,0.10,0.50)$ to $(0.20,0.20,1.00)$~mm, and the ECL position width from 10 to 20~mm.

In separate tests, we uniformly rescale all assigned shell material budgets to $s_{\text{mat}}=0.5$ or 1.5 and change $\epsilon_{\text{pair}}$ from 0.5 to 0.25 or 1.
For each material variation, we recalculate the conversion and survival probabilities in Eqs.~\eqref{eqn:supp_single_conversion_weight} and~\eqref{eqn:supp_conversion_branch} using Eq.~\eqref{eqn:supp_layered_conversion_probability}.
All other response parameters retain their nominal values.

\begin{figure}[H]
\centering
\includegraphics[width=0.99\linewidth]{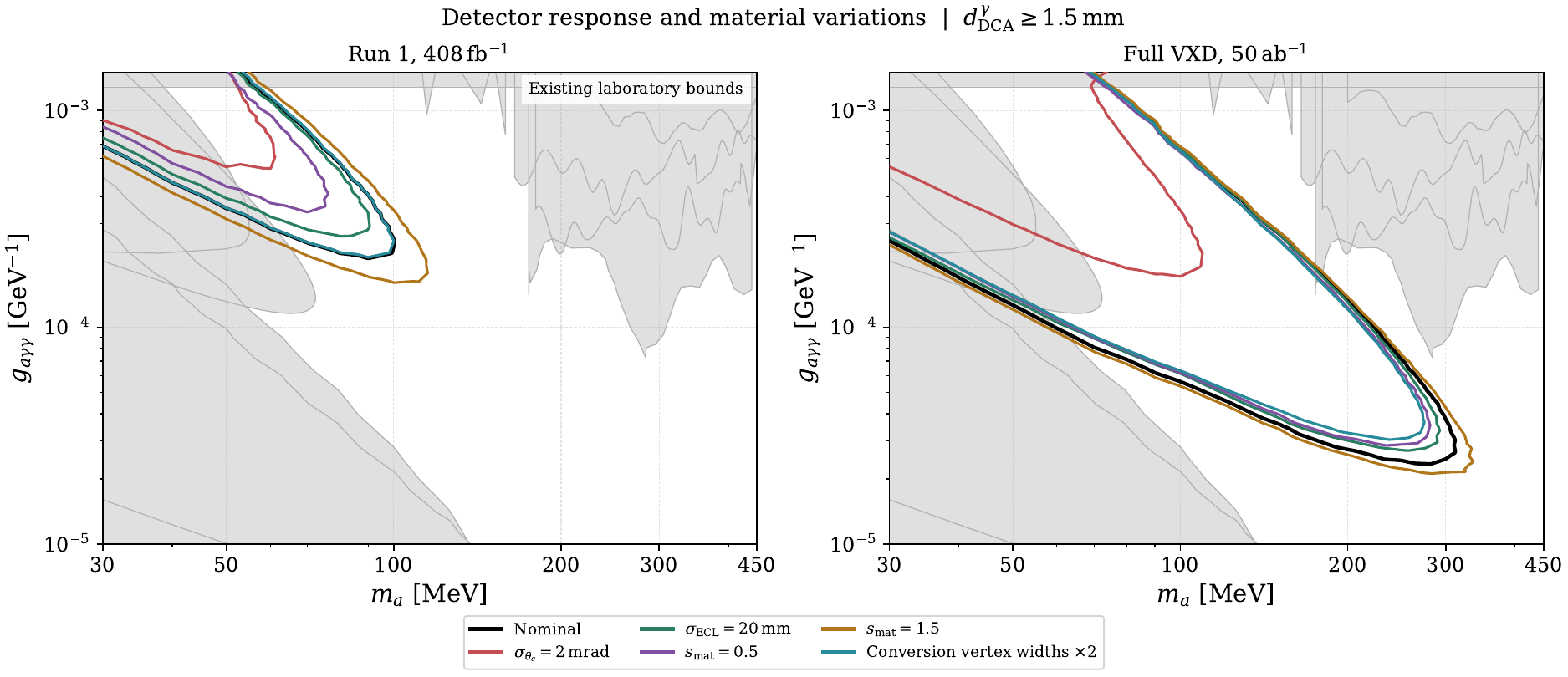}
\caption{Expected $95\%$ CL exclusion contours under individual detector-response and material variations, for $408$~fb$^{-1}$ with the Run~1 VXD (left) and $50$~ab$^{-1}$ with the full VXD (right). We keep the selection and $\epsilon_{\text{pair}}=0.5$ fixed and recalculate both signal and prompt-$3\gamma$ background. Shading shows existing laboratory exclusions.}
\label{fig:supp_robustness_detector}
\end{figure}

The converted-photon angular width has the largest effect on the projected sensitivity, as shown in Fig.~\ref{fig:supp_robustness_detector} and Table~\ref{tab:supp_robustness}.
At the tabulated benchmark, increasing $\sigma_{\theta_c}$ to 2~mrad changes the signal efficiency by only about $4\%$ but raises the selected prompt background by a factor of $\sim 200$.
Across the mass scan, this factor ranges from about 160 to 304.

At high luminosity, halving the shell material budgets reduces both signal and background rates, so the lower background partly compensates for the signal loss.
Doubling the conversion-vertex widths changes the benchmark signal efficiency by about $2\%$ but raises the background by a factor of 2.4.
At the same mass benchmark for $50$~ab$^{-1}$, $s_{95}$ is 8.57 for $s_{\text{mat}}=0.5$, 11.69 for the nominal model, and 17.41 for doubled conversion-vertex widths.
These changes explain why doubling the vertex widths weakens the high-luminosity reach more than halving the material budgets.
The ECL-position variation lowers both rates by changing reconstruction and selection acceptance.

For a constant pair-reconstruction efficiency, defining $\kappa_{\epsilon} \equiv \epsilon_{\text{pair}}/0.5$ gives $(s,B,\delta B_{\text{MC}})\to \kappa_{\epsilon}(s,B,\delta B_{\text{MC}})$.
We obtain the final two rows of Table~\ref{tab:supp_robustness} by recalculating the likelihood after this rate rescaling.

\begin{table}[H]
\centering
\small
\begin{tabular}{lrrrr}
\hline\hline
& & & \multicolumn{2}{c}{$g_{\min}/g_{\min}^{(0)}$}\\
Setting & $\epsilon_{\text{sig}}/\epsilon_{\text{sig}}^{(0)}$ & $B/B_0$ & $408~\text{fb}^{-1}$ & $50$~ab$^{-1}$\\
\hline
Nominal & 1.00 & 1.00 & 1.00 & 1.00\\
$\sigma_{\theta_c}=2$~mrad & 1.04 & 196.69 & 2.59 & 7.32\\
$\boldsymbol{\sigma}_{\text{v}}\times2$ & 0.98 & 2.37 & 1.01 & 1.30\\
$\sigma_{\text{ECL}}=20$~mm & 0.71 & 0.83 & 1.26 & 1.15\\
$s_{\text{mat}}=0.5$ & 0.52 & 0.54 & 1.63 & 1.22\\
$s_{\text{mat}}=1.5$ & 1.43 & 1.39 & 0.77 & 0.90\\
\hline
$\epsilon_{\text{pair}}=0.25$ & 0.50 & 0.50 & 1.69 & 1.22\\
$\epsilon_{\text{pair}}=1$ & 2.00 & 2.00 & 0.60 & 0.84\\
\hline\hline
\end{tabular}
\caption{Changes relative to the nominal model. The efficiency ratios use Run~1 at $m_a=100$~MeV and $c\tau_a=1.867$~mm, the lifetime-grid point nearest to 2~mm. The background ratios refer to the $m_{\text{hyp}}=100$~MeV window. The last two columns compare the lowest coupling on each contour within the ranges displayed in Fig.~\ref{fig:supp_robustness_detector}. These minima need not occur at the same mass. The nominal minima are $g_{\min}^{(0)}=2.08\times10^{-4}$ and $2.35\times10^{-5}$~GeV$^{-1}$ for $408$~fb$^{-1}$ and $50$~ab$^{-1}$, respectively. We obtain the final two rows by rescaling the nominal rates.}
\label{tab:supp_robustness}
\end{table}

\subsection{Residual-background stress test}
We test the sensitivity to additional backgrounds by keeping the nominal signal and rescaling the background as
\begin{equation}
B\to\kappa_B B,\qquad \delta B_{\text{MC}}\to\kappa_B\delta B_{\text{MC}},\qquad \text{with }\kappa_B=1,\,10,\,100.
\label{eqn:supp_background_stress}
\end{equation}
We apply the same $\kappa_B$ at every mass hypothesis, preserving the shape of $B_{3\gamma}(m_{\text{hyp}})$ and the relative MC uncertainty $\delta B_{\text{MC}}/B$.
Thus $N_{\text{eff},B}$ remains fixed while the absolute background rate and uncertainty increase.
For each $\kappa_B$, we recompute $s_{95}$ and the exclusion contour using the discrete-Poisson or asymptotic prescription appropriate to the rescaled background yield.

\begin{figure}[H]
\centering
\includegraphics[width=0.99\linewidth]{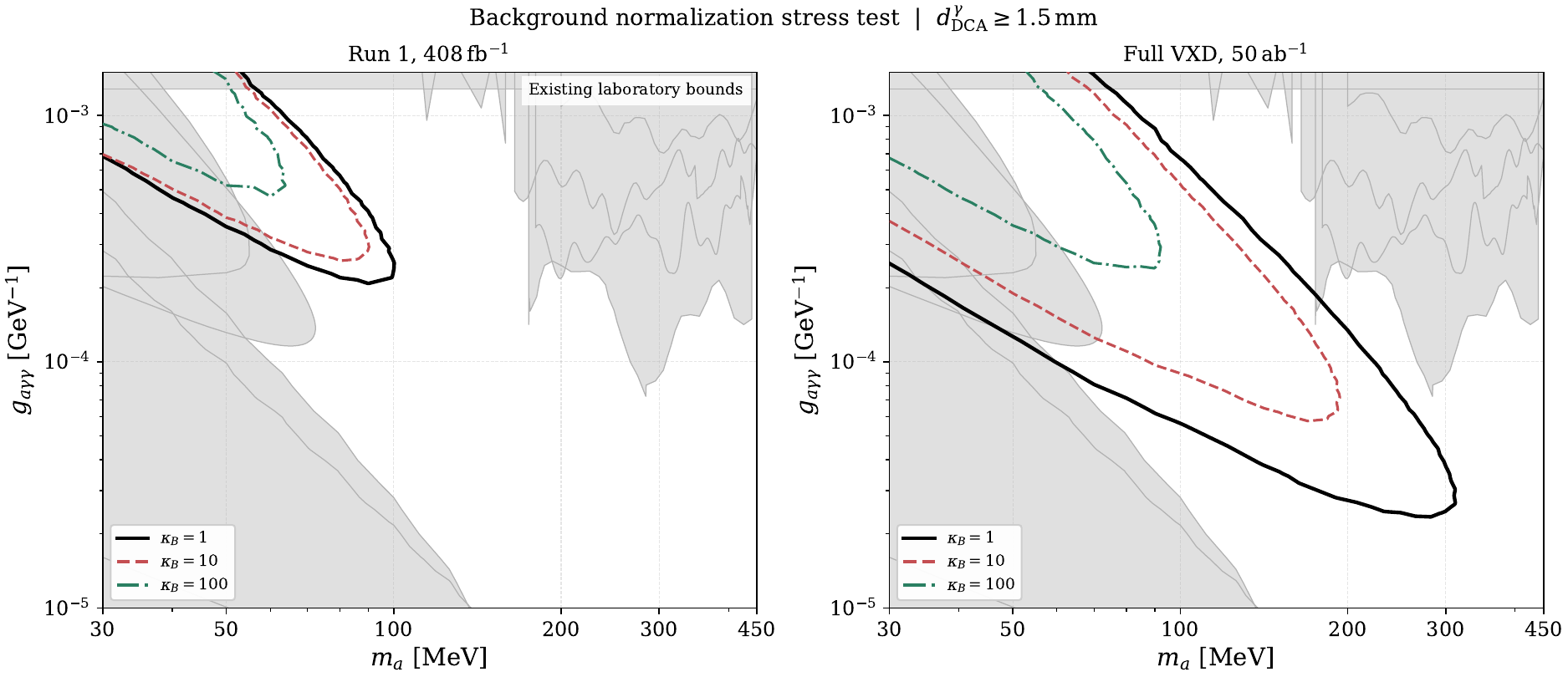}
\caption{Expected $95\%$ CL exclusion contours for the nominal detector response with $\kappa_B=1$, 10, and 100, using Eq.~\eqref{eqn:supp_background_stress}. We scale both the prompt-background rate and its MC standard error while keeping the signal and all selections fixed. The two panels and the existing exclusions follow Fig.~\ref{fig:supp_robustness_detector}.}
\label{fig:supp_robustness_background}
\end{figure}

Increasing the assumed background reduces the maximum mass reach, as shown in Fig.~\ref{fig:supp_robustness_background}.
For $408$~fb$^{-1}$, increasing $\kappa_B$ from 1 to 10 and 100 reduces the maximum mass within the displayed domain from about 100 to 90 and 64~MeV.
For $50$~ab$^{-1}$, it decreases from about 310 to 194 and 92~MeV.
Even the $\kappa_B=100$ scenario retains sensitivity to part of the parameter space outside the displayed existing laboratory exclusions.

\end{document}